\documentclass[12pt, draftclsnofoot,onecolumn]{IEEEtran}
\usepackage{dsfont}
\usepackage{amssymb}
\usepackage{float}
\usepackage{subfig}
\usepackage{graphicx, amssymb, amsmath}
\usepackage{extarrows}
\usepackage{algorithm} 
\usepackage{algorithmic} 
\usepackage{multirow} 
\usepackage{color}
\usepackage{cite}
\emergencystretch=\maxdimen
\IEEEoverridecommandlockouts
\begin{document}

\title{Hierarchical Soft Slicing to Meet Multi-Dimensional QoS Demand in Cache-Enabled Vehicular Networks}

\author{Shan~Zhang,~\IEEEmembership{Member,~IEEE,}
	Hongbin~Luo,~\IEEEmembership{Senior~Member,~IEEE,}
	Junling~Li,~\IEEEmembership{Student~Member,~IEEE,}
	Weisen~Shi,~\IEEEmembership{Student~Member,~IEEE,}
	and~Xuemin~(Sherman)~Shen,~\IEEEmembership{Fellow,~IEEE}
	\thanks{Shan~Zhang, and Hongbin~Luo are with Beijing Key Laboratory of Computer Networks, the School of Computer Science and Engineering, Beihang University, Beijing, China (email: \{zhangshan18, luohb\}@buaa.edu.cn).}
	\thanks{Junling~Li, Weisen~Shi, and Xuemin~(Sherman)~Shen are with the Department of Electrical and Computer Engineering, University of Waterloo, 200 University Avenue West, Waterloo, Ontario, Canada, N2L 3G1 (email: j742li, w46shi, sshen@uwaterloo.ca).}
}

\maketitle

\begin{abstract}
	
		Vehicular networks are expected to support diverse content applications with multi-dimensional quality of service (QoS) requirements, which cannot be realized by the conventional one-fit-all network management method. In this paper, a service-oriented hierarchical soft slicing framework is proposed for the cache-enabled vehicular networks, where each slice supports one service and the resources are logically isolated but opportunistically reused to exploit the multiplexing gain. The performance of the proposed framework is studied in an analytical way considering two typical on-road content services, i.e., the time-critical driving related context information service (CIS) and the bandwidth-consuming infotainment service (IS). Two network slices are constructed to support the CIS and IS, respectively, where the resource is opportunistic reused at both intra- and inter-slice levels. Specifically, the throughput of the IS slice, the content freshness (i.e., age of information) and delay performances of the CIS slice are analyzed theoretically, whereby the multiplexing gain of soft slicing is obtained. Extensive simulations are conducted on the OMNeT++ and MATLAB platforms to validate the analytical results. Numerical results show that the proposed soft slicing method can enhance the IS throughput by 30\% while guaranteeing the same level of CIS content freshness and service delay. 
	
\end{abstract}


\section{Introduction}
	
	Vehicular communication networks are expected to play a critical role in the future transportation systems, especially for intelligent driving assistance and travel experience enhancement.
	In addition, mobile edge caching will be leveraged to effectively support content services in this highly dynamic driving environment \cite{Amadeo16_ICVN_survey_mag}.
	In specific, the replica of popular contents can be pro-actively stored at the roadside units (RSUs) or on vehicles, whereby vehicles can obtain the requested contents in proximity, bringing three-fold benefits of shorter end-to-end delay, better mobility support, and reduced communication load \cite{Chang18_collaborative_mobile_clouds_mag,Su18_content_distribution_TVT}.
	Therefore, the cache-enabled vehicular networks are expected to be one of the supporting technologies in the intelligent transportation systems \cite{mine_VeCache_COMMAG,Guo18_video_cache_VNET_TVT}.
	
	Compared with conventional cellular networks, vehicular networks will accommodate more diversified applications with distinct features and quality of service (QoS) requirements \cite{mine_mag_auto_VNET}.
	For example, the infotainment service (IS), such as news, media and social entertainments, can bring enjoyable travel experiences to both drivers and passengers.
	In practice, IS is usually bandwidth-consuming and requires high network throughput.
	The context-related information service (CIS) is needed for intelligent driving assistance, such as the traffic flow of each road, the mobility of surrounding vehicles, the availability of parking lots, and the distance to next highway entrance.
	Unlike the IS, the CIS is usually time-critical, and the content information of a CIS item may change with time due to the variation of driving environment, calling for fresh content delivery within short delay.
	These distinct features pose great challenges to the management of cache-enabled vehicular networks.
	A fundamental issue is how to effectively meet the multi-dimensional QoS demands of different content services through efficient utilization of the heterogeneous resources of both RSUs and vehicles.	
	Existing studies have devoted extensive efforts on the QoS provisioning for IS and CIS, respectively. 
	However, the distinct features, multi-dimensional QoS demands, and the interplay relationships between different services are still lack of considerations.
	
	Network slicing is a promising paradigm to support differentiated services, whereby multiple network slices are constructed on-top-of a common network infrastructure to support multiple specified services.
	Each slice can work independently with logically isolated resources to meet the required QoS demands.
	Recently, slice-based vehicular network architectures have also been proposed to support different on-road applications \cite{Khan18_V2X_slicing_conf}.
	However, the detailed design of network slice management remains an open issue due to the challenges of resource efficiency, isolation guarantee, and multi-dimensional QoS requirements \cite{Campolo17_vehicular_slicing_mag}.
	
	In this work, we propose a hierarchical soft slicing framework for the cache-enabled vehicular networks to meet the multi-dimensional QoS demands of differentiated on-road content services.
	The typical CIS and IS are considered, and two service-oriented slices are constructed correspondingly.
	In the CIS slice, the RSUs cache the context information generated by distributed publisher nodes, and deliver the cached content items to vehicles on demand.
	Accordingly, the CIS slice consists of two functional blocks, i.e., the cache update function to guarantee the content freshness, and the content delivery function to transmit the requested contents to vehicles.
	In the IS slice, the infotainment content items are pro-actively pushed to vehicles through RSU broadcast, such that vehicles can be self-served or assisted through vehicle-to-vehicle (V2V) content sharing in proximity, i.e., local breakout.
	Accordingly, the IS slice also consists of two function blocks from the RSU aspect, i.e., the broadcast function to push newly generated popular contents to vehicles prior to requests, and the unicast function to serve content-miss vehicles as compensation.
	
	Under the proposed framework, the RSU transmission resource is shared at both intra- and inter-slice levels, and opportunistic resource reuse is introduced to enhance the system-level performance by exploiting the multiplexing gain.
	The key design issue is the intra- and inter-slice resource allocation, which affects the multi-dimensional QoS performances of both slices.
	To address this issue, the performances of both slices are analyzed.
	For the CIS slice, age of information (AoI), i.e., the time elapsed since the generation of the content, is adopted as the freshness metric.
	Both average AoI and delay are derived in closed-form expressions based on the random process analysis and queueing theory.
	For the IS slice, the throughput of local breakout is obtained by analyzing the proactive content pushing and V2V content sharing, with respect to the given communication and cache resources.
	Then, the total throughput of the IS slice is derived, demonstrating the coupling effect of intra- and inter-slice resource sharing with soft slicing.
	Based on the analytical results, the three-dimensional QoS performances of both slices are obtained, showing the multiplexing gain of soft slicing.
	Extensive simulations are conducted on the OMNeT++ and MATLAB platforms, validating the obtained analytical results.
	Numerical results are provided to show the effectiveness of proposed hierarchical soft slicing framework.
	In specific, the soft slicing method is shown to improve the IS throughput by around 30\% while maintaining the same QoS of the CIS slice, in comparison with the hard slicing scheme where the resources are shared orthogonally.
	Furthermore, the achievable QoS performance is illustrated under the soft slicing framework, which demonstrates a three-dimensional tradeoff relationship among the CIS content freshness, service latency, and IS throughput.
	This coupling effect on inter- and intra-slice QoS performance provides insights into practical resource management.
	The main contributions of this paper are:
	\begin{enumerate}
		\item A hierarchical soft slicing framework has been proposed to support the multi-dimensional QoS demands raised by differentiated content services in the cache-enabled vehicular networks, where resources are orthogonally allocated but opportunistically reused to guarantee logical isolation with enhanced efficiency at both inter- and intra-slice levels;
		\item The process of content update and transmission is analyzed in the CIS slice. The average AoI and delay are derived, which shows a tradeoff relationship with respect to intra-slice resource allocation.
		\item For the IS slice, the throughput of local breakout is obtained by analyzing the pro-actively content pushing and sharing. In addition, the total achievable throughput is derived with inter- and intra-slice resource reuse.
		\item The multi-dimensional QoS performance of the proposed slicing framework is evaluated with optimized inter- and intra-slice resource allocation. Numerical results show that the proposed hierarchical soft slicing framework can improve the IS throughput by 30\% compared with the optimized hard slicing method.
	\end{enumerate}

	The remaining of this paper is organized as follows. 
	Section~\ref{sec_review} presents a review of related works on content services of vehicular networks and network slicing.
	Section~\ref{sec_framework} introduces the proposed hierarchical soft slicing framework, followed by the system model developed for analysis in Section~\ref{sec_system_model}.
	Then, the performances of CIS and IS slices are analyzed in Sections~\ref{sec_CI_analysis} and \ref{sec_II_analysis}, respectively.
	Section~\ref{sec_simulation} conducts simulations and provides numerical results.
	Finally, conclusions are drawn in Section~\ref{sec_conclusions}.

\section{Literature Review}
	\label{sec_review}
	
	This section reviews existing works on cache-enabled vehicular network content services, according to the considered application types. 
	
	\subsection{Infotainment Service}
	
	Regarding the IS, the vehicular network performance has been analyzed theoretically under different communication modes \cite{Chetlur18_VANET_coverage_TWC} in addition to the real-trace evaluations \cite{Cao18_VNET_evaluation_real_trace_TVT,Sepulcre17_VNET_6mbps_measurement_TMC}.
	To further enhance the performance of IS, different methodologies have also been proposed, from different aspects such as data dissemination \cite{Dai18_VANET_dissemination_TVT}, cooperative transmission \cite{Chen18_coop_VNet_TVT}, and resource allocation \cite{Liang18_VNET_RA_TWC,Li17_VNET_channel_allocation_TVT_LianZhao}.
	In specific, pushing IS contents to vehicles prior to requests enables local traffic breakout, whereby vehicles can obtain the required contents from the on-board cache or nearby vehicles, reducing the burden of RSUs \cite{Hu19_cache_on_board_TVT}.
	Following this lead, on-board caching schemes have been devised to enhance the local content hit rate, through mobility-based caching \cite{Yao18_vehicle_cache_mobility_TVT}, social-aware caching \cite{Khan16_ICVN_social_vehicle_rank_mag}, and cache instance sharing with crowdsourcing \cite{Li18_ICVN_content_distri_crowds_Access}.
	As the popularity of an IS content item may fade with time, cache update is employed to maintain the hit rate.
	To this end, existing works have devised popularity-based content caching and update schemes for vehicles, where the vehicle-cached contents are dynamically replaced based on the prediction of popularity variations \cite{Zhao17_ICVN_popul_replace_Access}.
	However, cache updating also consumes transmission resources, which is seldom analyzed in the existing literature.
	
	\subsection{Context Information Service}
	
	The research on CIS has mainly aimed to reduce the service latency or enhance the reliability, by improving routing and access protocols or exploiting vehicular mobility information \cite{Katsaros16_VNET_delay_analysis_TVT,Lyu18_MoMac_TVT,Gao18_safety_message_broadcast_TVT}.
	Furthermore, the content information of a CIS item may change with time, calling for timely update to guarantee that vehicles receive fresh and effective information.
	AoI has been proposed in vehicular communication environment as a performance metric to depict content freshness, which is now getting increasing attentions \cite{Kaul11_AOI_concept_conf}. 
	To effectively support the time-critical CIS applications, both AoI and service latency should be guaranteed, such that vehicles can rapidly obtain the fresh contents on demand.
	However, these two objectives may not be consistent in practice, since frequent content updates improves content freshness but consumes additional communication resources \cite{Kaul12_AoI_basic_update_infocom}.
	Yet, there has been no study that jointly considers AoI and latency performances in vehicular networks, according to the authors' knowledge.

	\subsection{Coexistence of Differentiated Services}
	
	The study on differentiated services is still at the infant age in vehicular networking, as most related studies focus on a single service type.
	The very recent work in \cite{Van18_Vcaching_diff_app_IET} has proposed an application-aware content placement framework under the content-centric vehicular network architecture, whereby each vehicle determines which content to cache based on the attributes of the content as well as the mobility status and available on-board cache resource.
	Although insightful, \cite{Van18_Vcaching_diff_app_IET} mainly provides heuristic design, which may fail to guarantee the QoS performance in practice.
	
	The key enabling technology, network slicing, has been recently introduced into vehicular networks to support differentiated services \cite{Khan18_V2X_slicing_conf,Campolo18_V2X_slicing_conf}.
	However, there is still a lack of detailed designs on the management of heterogeneous vehicular network resources to guarantee the requirements of different slices.
	Meanwhile, the existing studies on cellular network slicing cannot capture the dynamics of vehicles, which makes it difficult to be directly applied to the vehicular networking scenarios \cite{junling_survey,junling_towardEnforcing,Ye18_slicing_hetnet_TVT}.
	
	Our previous work in \cite{mine_JSAC_VNET} has presented a slice-based management framework to support differentiated vehicular applications of map and navigation, file downloading, and on-demand service, through the collaborations of high altitude platform broadcast and RSU transmissions.
	The main differences between this work and \cite{mine_JSAC_VNET} are three-fold, i.e., scenario, content service types, and slicing methods.
	Firstly, this work considers the cooperation between cache-enabled RSUs and vehicles without the help of HAPs.
	Secondly, this work provides the distinct context information and infotainment services with multi-dimensional QoS requirements, whereas the services considered in \cite{mine_JSAC_VNET} mainly belong to the infotainment service.
	Thirdly, the resource is shared orthogonally with isolation across slices in \cite{mine_JSAC_VNET} and most existing research works, whereas this work enables opportunistic resource sharing in a soft slicing manner. This brings more challenges to QoS performance analysis and guarantee due to the intra- and inter-slice coupling effects.

\section{Hierarchical Soft Slicing Framework}
	\label{sec_framework}
	
	\begin{figure}[!t]
		\centering
		\includegraphics[width=3in]{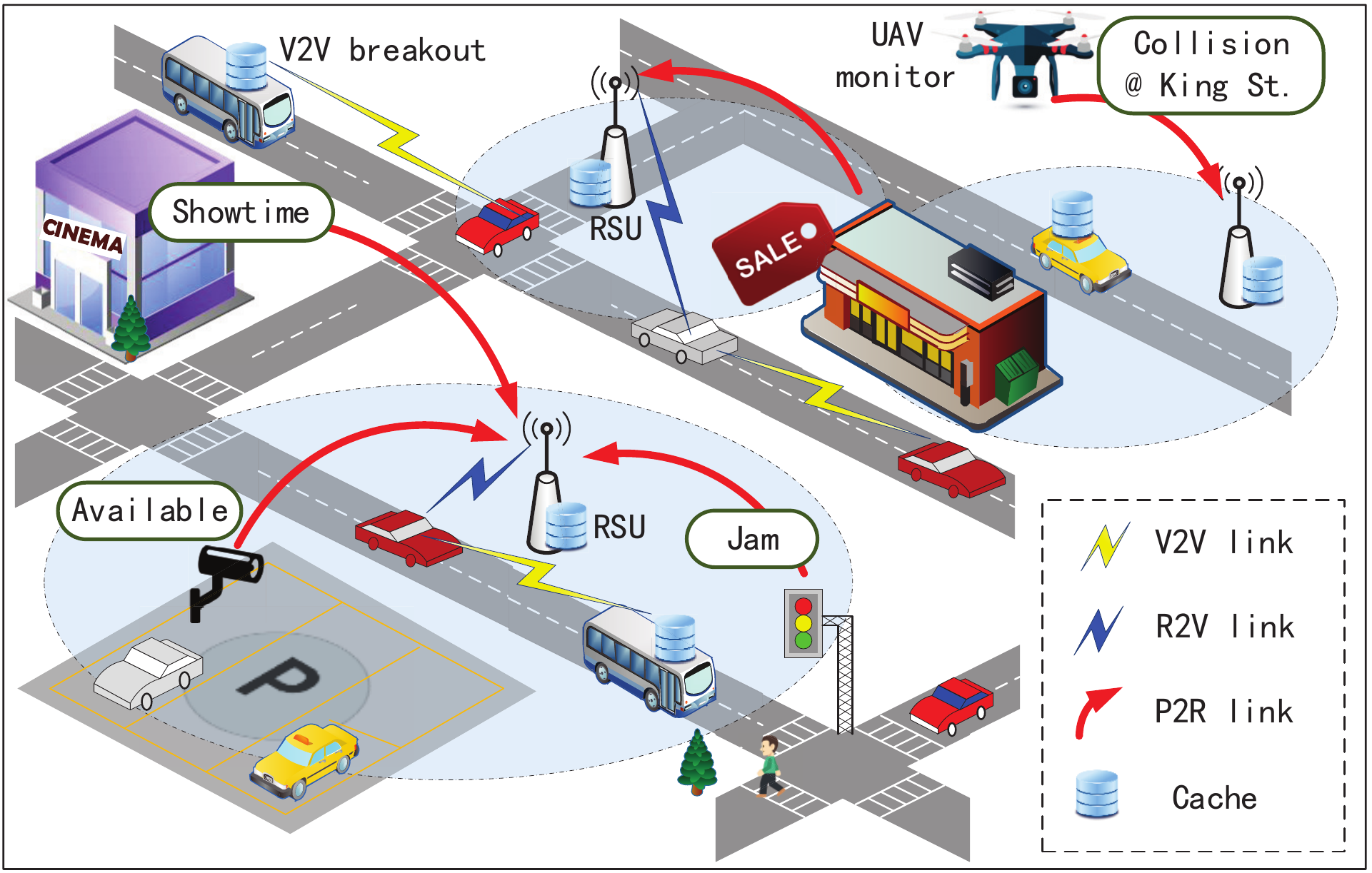}
		\caption{Illustration of cache-enabled vehicular networks.}
		\label{fig_scenario}
	\end{figure}
	
	\subsection{Vehicular Network Scenario}
	
	We consider a typical vehicular network providing both context information and infotainment services, as shown in Fig.~\ref{fig_scenario}.
	Three types of nodes are consisted: (1) \textbf{publishers} generating content items, (2) \textbf{vehicles} requiring content services, and (3) \textbf{RSUs} collecting and delivering content items.
	All the RSUs and part of the vehicles are equipped with cache instances.
	Three wireless communication modes are adopted: (1) publisher to RSU (P2R), whereby the publishers transmit contents to RSUs, (2) RSU to vehicles (R2V), whereby the RSUs deliver the requested contents to vehicles, and (3) vehicles to vehicles (V2V), whereby vehicles share cached contents in proximity to reduce the load of RSUs.
	
	For the CIS, the distributed publishers generate content items such as the availability of the parking lot, the intersection passing time, the promotion of nearby shopping malls, and the showtime of a cinema, as shown in Fig.~\ref{fig_scenario}.
	The RSUs work as information sinks to collect and cache the content items which may be requested by vehicles.
	Note that the information of a context content item may change with time due to the environment dynamics, and thus the publishers keep generating new content versions to reflect the instant status.
	Meanwhile, the RSUs fetch the newest versions for cache update, to avoid sending out-dated information to vehicles.
	When a vehicle raises a CIS request, the associated RSU will deliver the content item directly, without fetching the content from publishers.
	
\begin{figure}[!t]
		\centering
		\includegraphics[width=2.2in]{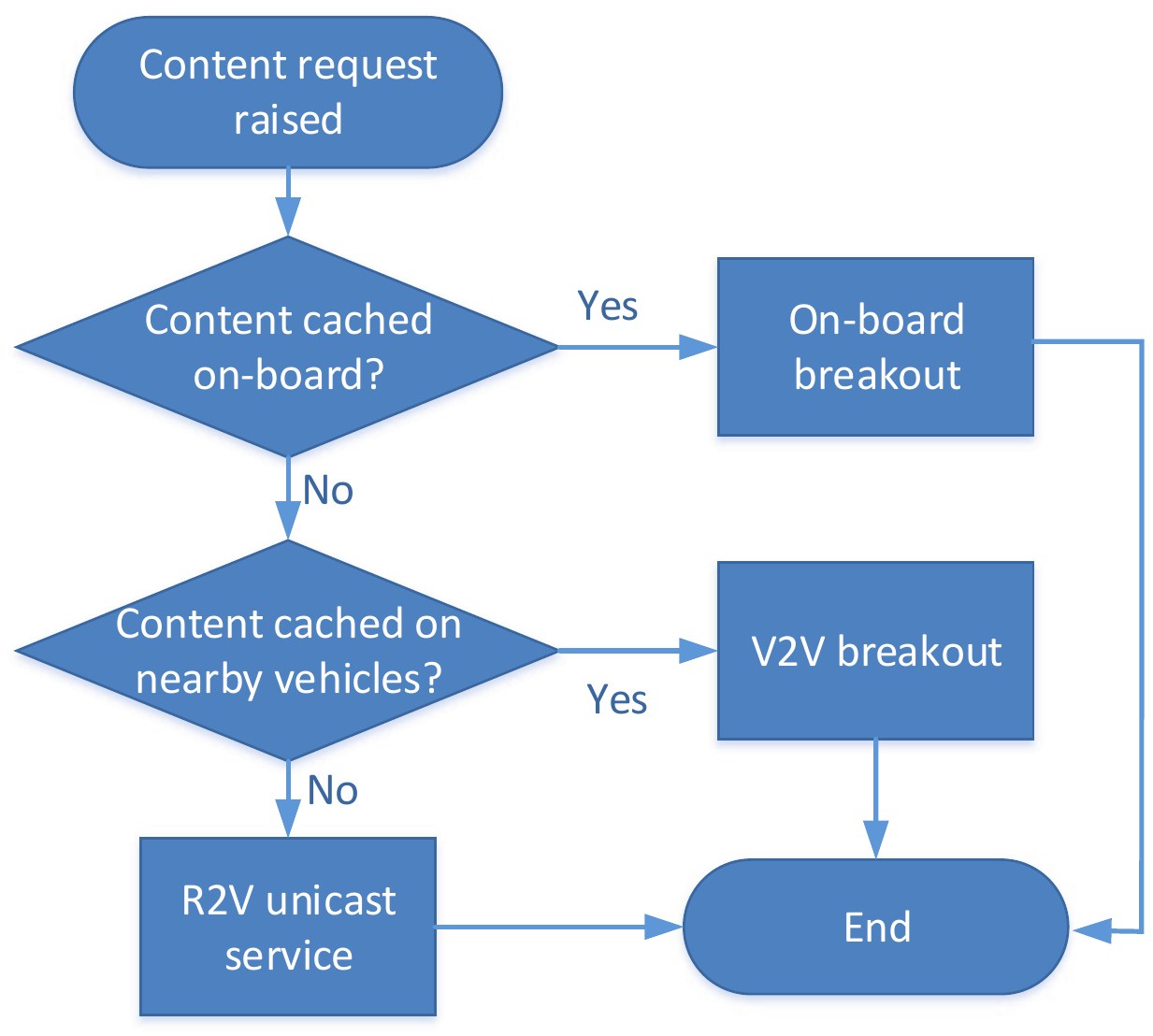}
		\caption{Flow chart to serve an infotainment content request.}
		\label{fig_service_process}
	\end{figure}

	Different from the CIS content items, the content of an IS item can be considered as static or semi-static, which seldom changes in small time scales.
	Therefore, the IS items can be cached on vehicles in prior to requests.
	In addition, vehicles can share contents in proximity through V2V transmissions to further reduce the load of RSUs.
	To this end, the RSUs push the popular IS contents to the vehicles through broadcast, which is highly transmission-efficient.
	The service process of the IS is shown as Fig.~\ref{fig_service_process}.
	
	\subsection{Service-Oriented Network Slicing}
	
\begin{figure}[!t]
		\centering
		\includegraphics[width=2.8in]{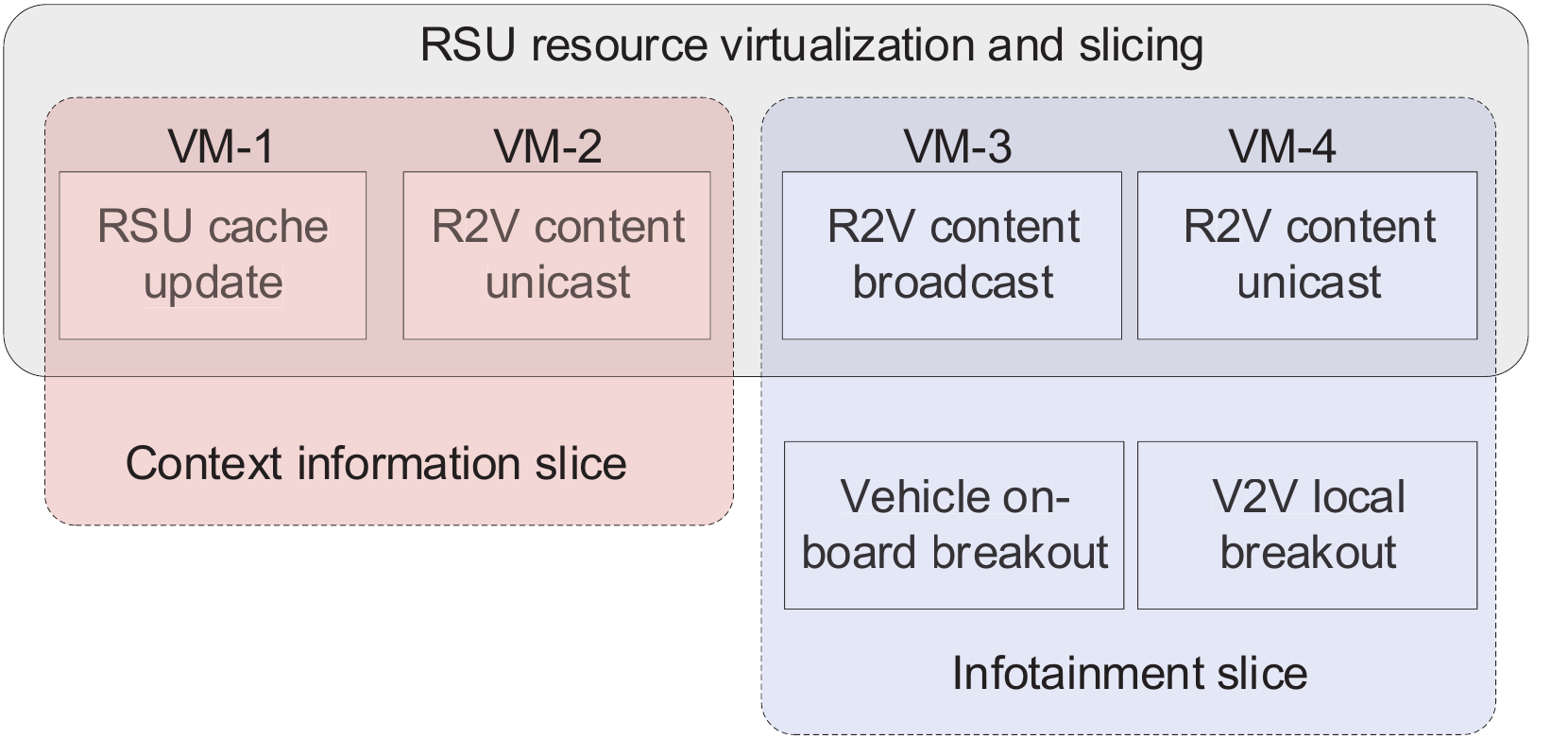}
		\caption{Illustration of the hierarchical slicing framework.}
		\label{fig_slicing_framework}
	\end{figure}

	To provide satisfying CIS, the RSUs should deliver timely information to vehicles with rapid transmissions.
	In contrast, the infotainment service is bandwidth-consuming but non-time-critical, which can endure service delay and content staleness.
	Considering these distinct features, a service-oriented network slicing framework is employed to satisfy the differentiated demands efficiently.
	In specific, each slice is responsible for one service, consisting of organized functional blocks, as shown in Fig.~\ref{fig_slicing_framework}\footnote{The framework can support more types of applications by constructing multiple service slices.}.
	Two functions are embedded in the CIS slice, both realized at RSUs using the wireless transmission resources.
	The RSU cache update function enables RSUs to fetch new versions of cached content items from publishers.
	The R2V content unicast function is responsible to deliver the requested content to vehicles through unicast.
	The IS slice consists of four functional blocks, according to the service process of Fig.~\ref{fig_service_process}.
	With the on-board cache function of vehicles, the R2V content broadcast function pushes contents to vehicles pro-actively.
	Then, vehicles requesting infotainment contents can be serviced locally through the on-board or V2V breakout functions.
	In addition, the content-miss vehicles are served through the R2V unicast transmission.
	
\begin{figure}[!t]
		\centering
		\includegraphics[width=2.5in]{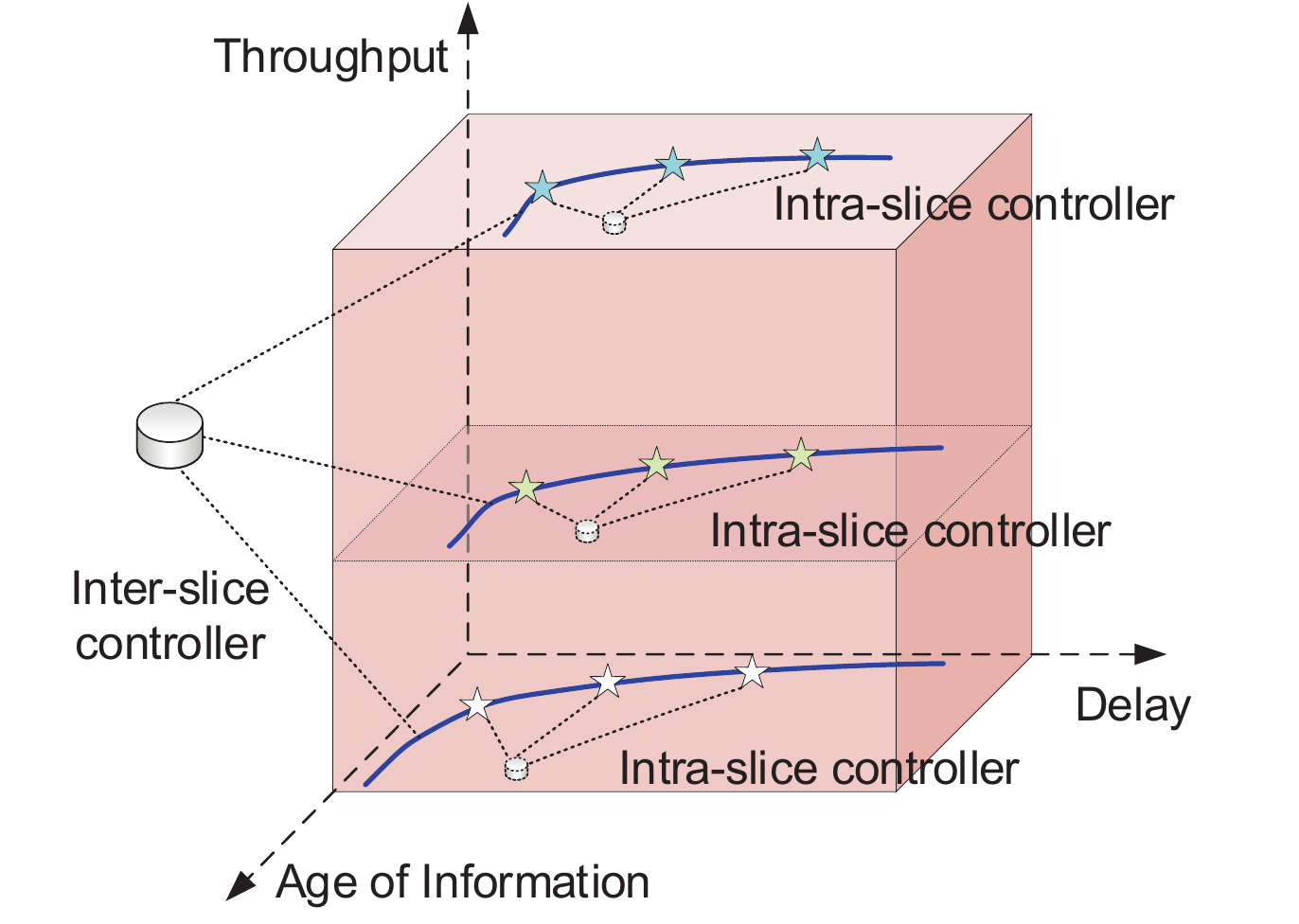}
		\caption{The multi-dimensional QoS metrics with respect to hirarchical soft network slicing.}
		\label{fig_QoS}
	\end{figure}

	Notice that the RSU transmission resources are shared at the inter- and intra-slice levels to support all the required functions.
	Furthermore, the cache resource and V2V links are integrated in the IS slices to enhance the service capability. 
	The detailed hierarchical slicing schemes influence the QoS performance of both slices due to the constrained resources.
	Figure~\ref{fig_QoS} illustrates the coupling effects of the QoS metrics, where the delay and AoI reflect the performance of the CIS slice while the IS slice pursues throughput.
	The solid lines marked by stars depict the QoS performance under different inter- and intra-slicing ratios.
	As the inter-slice controller allocates more resources to the IS slice, the corresponding throughput increases, whereas the delay and AoI performance of the CIS slice degrades.
	Furthermore, the AoI and delay may show a tradeoff relationship for the given inter-slice resource allocation, since the RSU cache update and R2V content unicast functions share the RSU transmission resources.
	The tradeoff relationship can be tuned by the intra-slice controller which determines the resource sharing ratio.
	The hierarchical slicing should be carefully designed to achieve the optimal balance among the multi-dimensional QoS performance metrics, which is the main focus of this work.
	
\section{System Model}
    \label{sec_system_model}
    
    This section builds the system model of the vehicular network under the soft hierarchical slicing framework.
    
    \subsection{Network Topology Model}
    
    We consider a one-dimensional road consisting of $J$ unidirectional or bidirectional lanes.	
    The RSUs are deployed regularly beside the road with coverage radius of $R_\mathrm{R}$, and are	equipped with sufficient cache instances to cache all CIS content items.
    The context information publishers are randomly distributed in network following two-dimensional Poisson point process (PPP) of density $\rho_\mathrm{p}$.
    Denote by $\rho_\mathrm{v}$ the density of vehicles in each lane, which are also uniformly distributed following one-dimensional PPP.
    Part of the vehicles are equipped with cache instances of size $C_\mathrm{v}$, and the ratio of cache-enabled vehicles is $P_\mathrm{vc}$.
    
\begin{figure}[!t]
    	\centering
    	\includegraphics[width=2.8in]{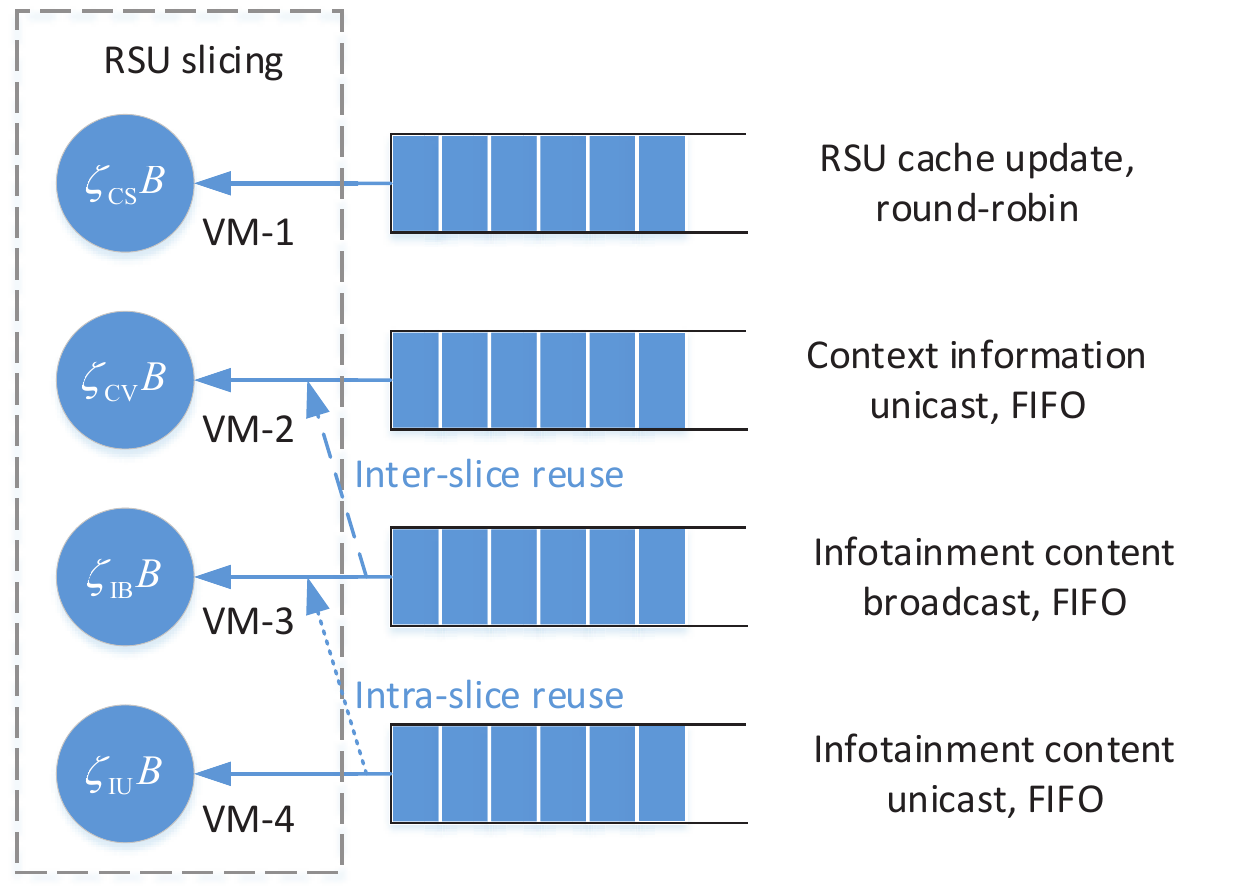}
    	\caption{The RSU service process model under the soft hierarchical network slicing framework.}
    	\label{fig_queue}
    \end{figure}

    The service process of the RSU can be modeled as a queueing system with processor sharing, as shown in Fig.~\ref{fig_queue}.
    The bandwidth resources are sliced and shared as four virtual machines (VMs), corresponding to the four functional blocks in Fig.~\ref{fig_slicing_framework}.
    Denote by $B$ the available bandwidth of the RSU, $\zeta_\mathrm{CS}$, $\zeta_\mathrm{CV}$, $\zeta_\mathrm{IB}$ and $\zeta_\mathrm{IU}$ the resource slicing ratio, where $\zeta_\mathrm{CS} + \zeta_\mathrm{CV} + \zeta_\mathrm{IB} + \zeta_\mathrm{IU} = 1$.
    
    VM-1 is utilized  for cache update, where all the cached CIS items are updated in a round-robin manner.
    VM-2 is responsible to deliver the requested CIS contents to vehicles through R2V unicast.
    The corresponding service requests are served in a first-in-first-out (FIFO) manner.
    VM-3 belongs to the IS slice, which is used by the RSUs to push the popular IS contents to vehicles through broadcast.
    VM-4 serves the vehicles which fail to obtain the requested IS content on-board or through V2V breakout.
    R2V unicast communication mode is used, and the requests are served in the FIFO way.
    With soft slicing, the VMs can be reused opportunistically by other functions to enhance the network efficiency, within a slice (i.e., intra-slice reuse) or across slices (i.e., inter-slice reuse), as shown by the dashed lines in Fig.~\ref{fig_queue}.
    VM-2 of the CIS slice can be reused to broadcast the popular IS contents, i.e., inter-slice reuse.
    Additionally, the VM-3 can be reused to serve the cache-miss vehicles through unicast, which is a way of intra-slice resource reuse.  
    Notice that the VMs are only reused in the idle state, i.e., the corresponding queue is empty.
    Therefore, each slice still enjoys logically isolated resource, and the QoS can be guaranteed without interference.

    \subsection{Wireless Traffic Model}

    Denote by $\mathcal{S}=\{1,2,\cdots,S\}$ the set of CIS content items to be requested, where $S=|\mathcal{S}|$.
    Denote by $\mathcal{F}=\{1,2,\cdots,F\}$ the file library of IS slice, where $F$ is the total number of files and $F\geq C_\mathrm{v}$ considering the constrained cache instances.
    Furthermore, the IS files may show differentiated popularity.
    Denote by $q_f$ ($f=1,2,...,F$) the probability that file $f$ is requested, where $\sum_{f=1}^{F} q_f = 1$.
    
    Continuous time model is adopted to analyze the request and service process. 
    For both slices, vehicles raise requests of content items randomly, following a Poisson process.
    Denote by $\lambda_\mathrm{C}$ and $\lambda_\mathrm{I}$ the request arrival rate per vehicle of the CIS and the IS slices, respectively.
    Thus, at the RSU, the aggregated request arrival process of the CIS slice also follows Poisson process, at rate of $\Lambda_\mathrm{C} = 2\pi J R_\mathrm{R} \rho_\mathrm{v} \lambda_\mathrm{C}$.
    However, the aggregated traffic load of the IS slice is reduced with local breakout.
    Denote by $P_\mathrm{Hit}$ the probability that the requested IS content is stored on a cache-enabled vehicles, and $P_\mathrm{V2V}$ the probability that a vehicle without cache can be assisted by surrounding cache-enabled vehicles.
    Thus, the local breakout probability of IS contents is given by
    \begin{equation}
    \label{eq_P_LB}
    P_\mathrm{LB} = P_\mathrm{Hit} \left(P_\mathrm{vc} + (1-P_\mathrm{vc})P_\mathrm{V2V} \right),
    \end{equation}
    where the two parts corresponding to on-board breakout and V2V-assisted breakout, respectively.
    Accordingly, the RSU aggregated traffic load of the IS slice is given by 	
    \begin{equation}
    \Lambda_\mathrm{I} =  2\pi J R_\mathrm{R} \rho_\mathrm{v} \lambda_\mathrm{I} (1-P_\mathrm{LB}).
    \end{equation}
    
    \subsection{Wireless Communication Model}
    
    To conduct theoretical analysis, we assume that the service time of each VM follows exponential distribution due to the random wireless channel conditions.
    Applying the Shannon formula, the average service rates of the four VMs can be obtained:
    \begin{equation}
    \begin{split}
    \mu_\mathrm{CS} & =\underset{d_\mathrm{RP}}{\mathds{E}}\left[\frac{\zeta_\mathrm{CS} B}{L} \log_2\left( 1+\frac{P_\mathrm{TP} d_\mathrm{RP}^{-\alpha}}{I_\mathrm{Intf}+\sigma^2} \right) \right] \triangleq \zeta_\mathrm{CS} \tilde{\mu}_\mathrm{CS},\\
    \mu_\mathrm{CV} & = \underset{d_\mathrm{RV}}{\mathds{E}} \left[\frac{\zeta_\mathrm{CV} B}{L} \log_2\left( 1+\frac{P_\mathrm{TR} d_\mathrm{RV}^{-\alpha}}{I_\mathrm{Intf}+\sigma^2} \right) \right] \triangleq \zeta_\mathrm{CV} \tilde{\mu}_\mathrm{CV}, \\
    \mu_\mathrm{IB} & = \frac{\zeta_\mathrm{IB} B}{L} \log_2\left( 1+\frac{P_\mathrm{TR} D_\mathrm{RV}^{-\alpha}}{I_\mathrm{Intf}+\sigma^2} \right) \triangleq \zeta_\mathrm{IB} \tilde{\mu}_\mathrm{IB},\\  			
    \mu_\mathrm{IU} & = \underset{d_\mathrm{RV}}{\mathds{E}}\left[\frac{\zeta_\mathrm{IU} B}{L} \log_2\left( 1+\frac{P_\mathrm{TR} d_\mathrm{RV}^{-\alpha}}{I_\mathrm{Intf}+\sigma^2} \right) \right] \triangleq \zeta_\mathrm{IU} \tilde{\mu}_\mathrm{IU},
    \end{split}
    \end{equation}
    where $L$ is the file size\footnote{The files are assumed to have the same size for the simplicity of analysis. In case of unequal file size, each file can be divided into multiple chunks of the same size to apply the proposed method.}, $\alpha$ is the path loss factor, $I_\mathrm{Intf}$ is the interference power, $\sigma^2$ is the addictive Gaussian noise power, $P_\mathrm{TP}$ (and $P_\mathrm{TR}$) is the transmit power of a publisher node (and a RSU), $d_\mathrm{RP}$ and $d_\mathrm{RV}$ are random variables representing the transmission distances under the P2R and the R2V unicast communication modes, $D_\mathrm{RV}$ is the maximal transmission distance under R2V broadcast mode, $\tilde{\mu}_\mathrm{CS}$, $\tilde{\mu}_\mathrm{CV}$, $\tilde{\mu}_\mathrm{IB}$, and $\tilde{\mu}_\mathrm{IU}$ denote the normalized average service rates of the four VMs when allocated with all bandwidth, respectively. The normalized average service rates can be obtained by approximated theoretical analysis or numerical calculation for the given system parameters \cite{mine_cache_TMC}.

    With inter-slice reuse, the IS content broadcast rate is further enhanced by
    \begin{equation}
    \label{eq_delta_ib}
    \Delta_\mathrm{IB} = \zeta_\mathrm{CV} \mu_\mathrm{IB} o_\mathrm{VM2},
    \end{equation}
    where $o_\mathrm{VM2}$ is the probability that VM-2 is in the idle state with no request to serve.
    The intra-slice reuse of VM-3 enhances the IS unicast rate by:
    \begin{equation}
    \Delta_\mathrm{IU} = \zeta_\mathrm{IB} \mu_\mathrm{IU} o_\mathrm{VM3},
    \end{equation}
    where $o_\mathrm{VM3}$ is the probability that VM-3 is in the idle state.
    
    The key issue of the soft hierarchical slicing is to analyze the QoS performance of each service and find the optimal resource allocation at both inter- and intra-slice levels.
    For the context information slice, the service rate of VM-1 mainly influences the freshness of contents cached at RSUs, whereas the service rate of VM-2 influences the service delay as well as the content freshness received by vehicles.
    Therefore, the inter- and intra-slice resource allocations jointly determine the freshness and delay performance of the CIS slice.
    This coupling effect is more significant for the IS slice.
    The inter-slice coupling effect exists as VM-2 can be reused to broadcast IS contents.
    For the intra-slice resource allocation, the service rate of VM-3 influences the probability of local breakout, while the service rate of VM-4 determines how much traffic the RSU can accommodate.
    Accordingly, the overall throughput of the IS slice is determined together by the service capabilities of VM-2, VM-3 and VM-4. 
    More detailed performance analysis will be conducted in the following sections.

\section{Context Information Slice Management}
    \label{sec_CI_analysis}
	
	In this section, the average AoI and delay of the CIS slice are derived by analyzing RSU cache update (i.e., VM-1) and content delivery (i.e., VM-2).
	Then, the optimal intra-slice resource allocation is obtained.
	
	\begin{figure}[!t]
		\centering
		\includegraphics[width=3in]{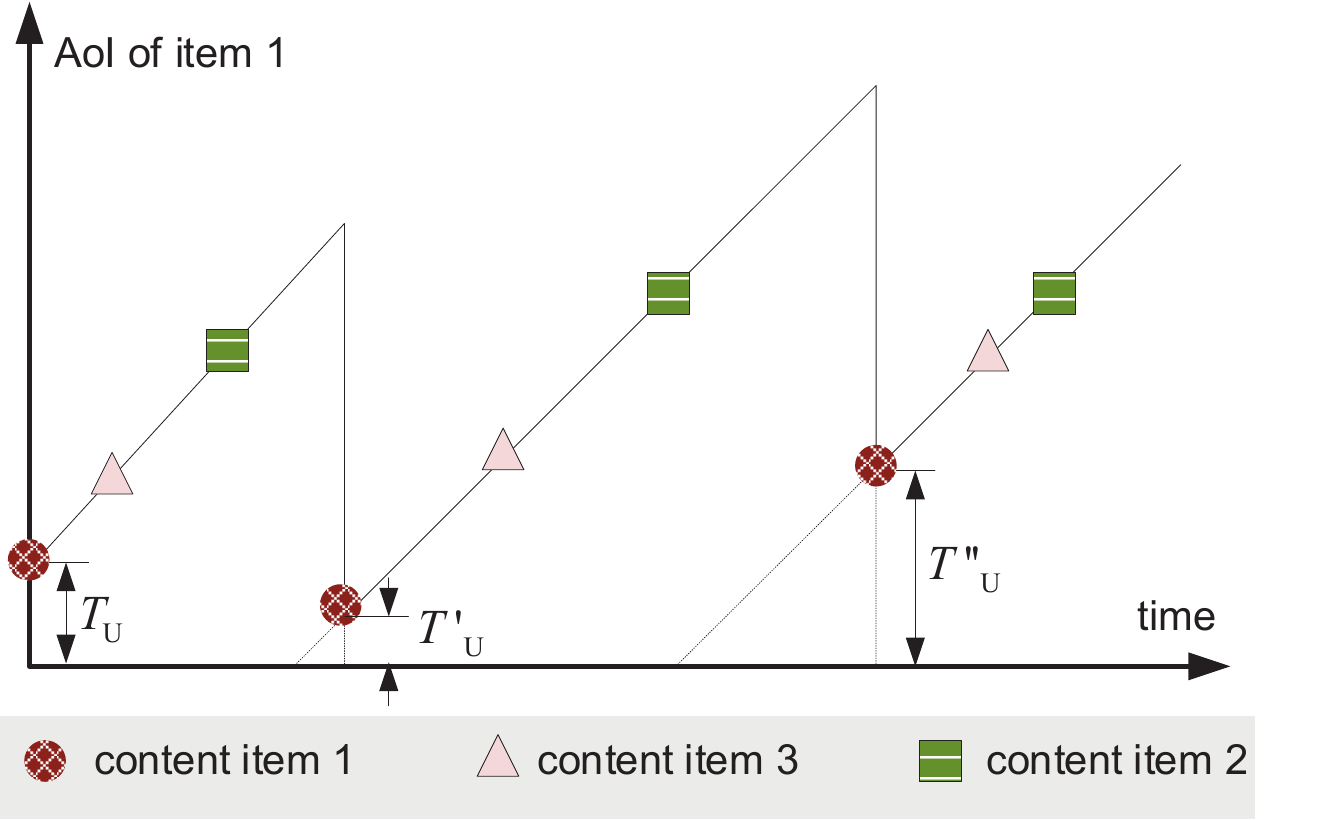}
		\caption{AoI variation illustration with RSU cache update.}
		\label{fig_AoI}
	\end{figure} 
	
	\subsection{RSU Cache Update}
	
	The RSU updates the cached CIS item in a round-robin manner utilizing VM-1, and the AoI of each content item varies with time.
	Figure~\ref{fig_AoI} illustrates the AoI variation in a three-item case, where content item 1 is updated at time zero.
	Once updated, the AoI of item 1 is reset to the transmission time from the publisher to the RSU, denoted as $T_\mathrm{U}$, $T'_\mathrm{U}$, and $T''_\mathrm{U}$.
	Then, the AoI increases linearly with time until the next update.
	
\begin{figure}[!t]
		\centering
		\includegraphics[width=2.2in]{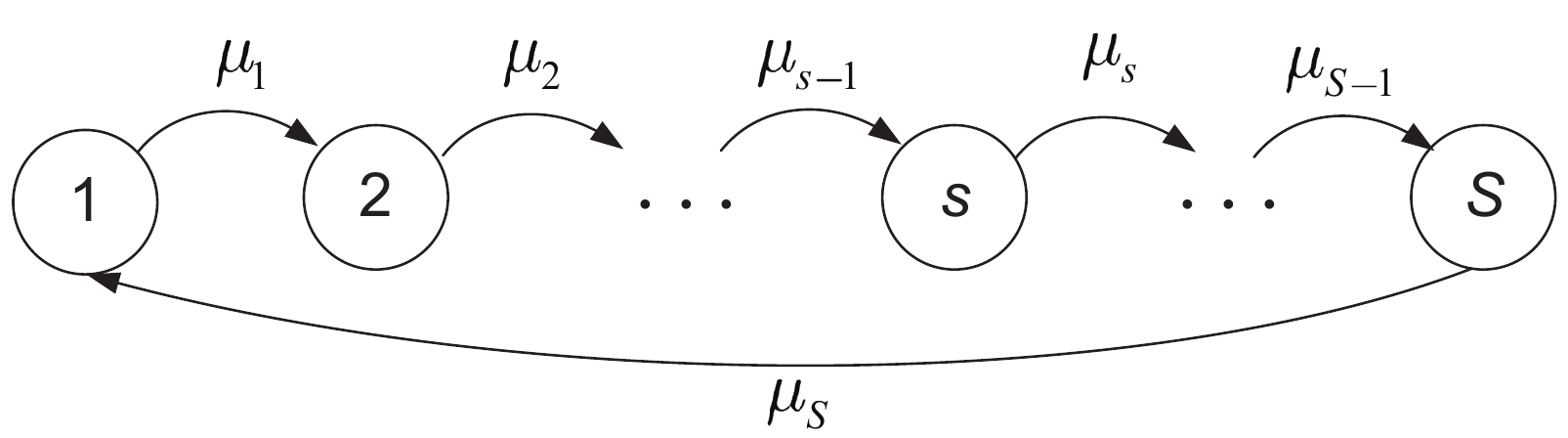}
		\caption{State transition of RSU cache update.}
		\label{fig_Mchain}
	\end{figure}

	This process can be analyzed by a Markov chain. 
	The set of state is denoted by $\mathbf{S} = [1,2,\cdots,S]$, where state $s \in \mathbf{S}$ means that the RSU is updating content item $s$.
	The process of state transition is shown as Fig.~\ref{fig_Mchain}, where $\mu_s$ ($s=1,2,...,S$) is the P2R transmission rate of content item $s$.
	In practical systems, $\mu_s$ can vary with publishers due to the different channel conditions.
	According to Fig.~\ref{fig_Mchain}, the state transition matrix of the Markov chain is given by
	\begin{equation}
	\mathbf{Q}_\mathrm{R}=
	\begin{bmatrix}
	-\mu_1     & \mu_1 & 0 &\cdots & 0      \\
	0 & -\mu_2 & \mu_2 & 0  & \vdots \\
	\vdots & \ddots & \ddots & \ddots & \vdots\\
	0 & \cdots & 0 & -\mu_{S-1} & \mu_{S-1} \\
	\mu_{S} & 0 &\cdots& 0&-\mu_{S}\\
	\end{bmatrix} .
	\end{equation}
	Denote by $\Psi=[\psi_1,\psi_2,...\psi_S]$ the stable state probability distribution, which should satisfy
	\begin{subequations}
		\label{eq_state_pi}
		\begin{align}
		& \Psi \mathbf{Q}_\mathrm{R} = \Psi,\\
		& \Psi \boldsymbol{e} = 1,
		\end{align}
	\end{subequations}
	where (\ref{eq_state_pi}) is the global balance function, $\boldsymbol{e}=[1,1,...,1]$ is a column vector of length $S$.
	Thus, the stable state probability distribution is given by:
	\begin{equation}
	\psi_s = \frac{\frac{1}{\mu_s}}{\sum_{i=1}^{S} \frac{1}{\mu_i}},~~~s=1,2,...,S.
	\end{equation}
	Notice that $\psi_s$ will increase if the update rate $\mu_s$ is small, which is reasonable as it takes longer time to complete the update in state $s$.	
	
	\subsection{Age of Information Analysis}
	
	Consider a vehicle requesting content item $s$.
	Suppose the content update Markov chain is in state $k$ when the RSU begins the delivery.
	Accordingly, the experienced states since the last update of content item $s$ is $\{s,s+1,...,k\}$ if $k>s$, and is $\{s,s+1,...,S, 1,2,...,k\}$, otherwise.
	Denote by $t_i$ the dwelling time in state $i$, which follows exponential distribution of mean $1/\mu_i$ for $i=1,2,...,S$.
	Accordingly, the AoI of content item $s$ at the RSU is given by
	\begin{equation}
	A_s^{\mathrm{(RSU)}} = \left\{ \begin{array}{ll}
	\sum_{i=s}^{k} t_i,&~~k>s,\\
	\sum_{i=s}^{S} t_i + \sum_{i=1}^{k} t_i,&~~k\leq s.
	\end{array} \right.
	\end{equation}
	Then, the average AoI of the RSU cached content item $s$ can be obtained:
	\begin{equation}
	\label{eq_AoI_RSU}
	\begin{split}
	& \bar{A}_s^{\mathrm{(RSU)}} =  \underset{\{k,\{t_i\}\}}{\mathds{E}} \left[A_s^{\mathrm{(RSU)}}\right],\\
	& = \underset{\{k,\{t_i\}\}}{\mathds{E}} \left[ \sum_{k=1}^{s} \psi_k \left(\sum_{i=s}^{S}t_i + \sum_{i=1}^{k}t_i\right) + \sum_{k=s+1}^{S}\psi_k \sum_{i=s}^{k}t_i \right]\\
	& = \frac{1}{\sum\limits_{i=1}^{S}\frac{1}{\mu_i}} \left[\sum_{k=1}^{s}\frac{1}{\mu_k}\left(\sum_{i=s}^{S}\frac{1}{\mu_i}+\sum_{i=1}^{k}\frac{1}{\mu_i}\right) + \sum_{k=s+1}^{S}\frac{1}{\mu_k}\sum_{i=s}^{k}\frac{1}{\mu_i}\right]\\
	& = \frac{1}{\sum\limits_{i=1}^{S}\frac{1}{\mu_i}} \left[ \sum_{k=1}^{s} \sum_{i=s}^{S}\frac{1}{\mu_k\mu_i} + \sum_{k=1}^{S}\sum_{i=1}^{k} \frac{1}{\mu_k\mu_i} + \sum_{k=s+1}^{S}\sum_{i=s}^{k} \frac{1}{\mu_k\mu_i}   \right]
	\end{split}
	\end{equation}
	We first consider the two-publisher case to analyze the influence of differentiated transmission rate of publishers.
	
	\textbf{Case 1. Two CIS publishers}
	
	Set $S=2$, $\mu_1 = \frac{1}{1+\epsilon} \mu$ and $\mu_2 = \frac{\epsilon}{1+\epsilon} \mu$, where $\mu$ is the sum transmission rate of the two publishers, and $\epsilon \leq 1$ reflecting the rate difference of the two publishers.
	The average AoI of content item 1 can be written as 
	\begin{equation}
	\label{eq_AoI_21}
	\begin{split}
	& \bar{A}_1^{\mathrm{(RSU)}} = \frac{\mu}{2+\epsilon+\frac{1}{\epsilon}}\left[ \frac{1}{\mu_1^2} + \frac{1}{\mu_1\mu_2} + \frac{1}{\mu_1^2} + \frac{1}{\mu_2^2} + \frac{1}{\mu_1\mu_2} \right] \\
	& = \frac{\epsilon}{(1+\epsilon)^2\mu} \left[ 2(1+\epsilon)^2+ 2(1+\epsilon)(1+\frac{1}{\epsilon}) + (1+\frac{1}{\epsilon})^2 \right]\\
	& = \frac{1}{\mu} \left(2\epsilon+2+\frac{1}{\epsilon}\right).
	\end{split}
	\end{equation}
	Similarly, the average AoI of content item 2 is given by
	\begin{equation}
	\label{eq_AoI_22}
	\bar{A}_2^{\mathrm{(RSU)}} = \frac{1}{\mu} \left(\epsilon+2+\frac{2}{\epsilon}\right).
	\end{equation}
	Thus, the average system-level AoI is
	\begin{equation}
	\bar{A}^{\mathrm{(RSU)}} \!=\! \frac{1}{2}\left(\bar{A}_1^{\mathrm{(RSU)}} \!+\! \bar{A}_2^{\mathrm{(RSU)}}\right) = \frac{1}{2\mu} \left(3\epsilon+4+\frac{3}{\epsilon}\right),
	\end{equation}
	which achieves the minimum when $\epsilon=1$.
	
	This case indicates that the symmetric networks show the optimal system-level AoI performance.
	Thus, we can approximate the transmission rate of each publisher by the average transmission rate (i.e., $\mu_s = \frac{\sum_{i=1}^{S}\mu_i}{S}$), which provides the lower bound of average AoI.
	
	\textbf{Case 2. Symmetric CIS publishers}
	
	Set $\mu_s=\bar{\mu}=\sum_{i=1}^{S}\mu_i/S$, the average AoI of each content item is equal, and (\ref{eq_AoI_RSU}) can be simplified as 
	\begin{equation}
	\label{eq_AoI_aver_S}
	\begin{split}
	\bar{A}^{\mathrm{(RSU)}} & = \frac{1}{S\bar{\mu}} \left[\sum_{k=1}^{s}\left(S-s+1+k\right) + \sum_{k=s+1}^{S}(k-S+1)\right]\\
	& =\frac{1}{S\bar{\mu}} \left[sS + \sum_{k=1}^{S}(k-S+1)\right]\\
	& = \frac{1}{S\bar{\mu}} (2+3+\cdots+S+1) = \frac{S+3}{2\bar{\mu}}.
	\end{split}
	\end{equation} 
	Then, the network-level AoI of user received CIS contents can be obtained approximately, given by Theorem~1.
	
	\textbf{Theorem~1.} For the CIS slice, the approximated network-level average AoI of user received contents is given by
	\begin{equation}
	\label{eq_A_v}
	\bar{A}^{\mathrm{(V)}} = \frac{\pi R_\mathrm{R}^2 \rho_\mathrm{p}+3}{2\zeta_\mathrm{CS}\tilde{\mu}_\mathrm{CS}} + \frac{1}{\zeta_\mathrm{CV}\tilde{\mu}_\mathrm{CV}}.
	\end{equation}
	
	\emph{Proof.}~From the network aspect, the number of publishers $S$ in each RSU varies randomly.
	As the publishers follows two-dimensional PPP of density $\rho_\mathrm{p}$, $S$ follows Poisson distribution of mean $\pi R_\mathrm{R}^2 \rho_\mathrm{p}$.
	Substituting into (\ref{eq_AoI_aver_S}), the network-level average AoI of RSU cached content items is given by 
	\begin{equation}
	\bar{A}^{\mathrm{(RSU)}} = \frac{\pi R_\mathrm{R}^2 \rho_\mathrm{p}+3}{2\zeta_\mathrm{CS}\tilde{\mu}_\mathrm{CS}}.
	\end{equation}
	Furthermore, the AoI of user received contents includes the staleness introduced during content delivery:
	\begin{equation}
	\bar{A}^{\mathrm{(V)}} = \bar{A}^{\mathrm{(RSU)}} + \mathds{E}(T_\mathrm{D}),  
	\end{equation}
	where $T_\mathrm{D}$ is the transmission time from RSU to vehicle, following exponential distribution of $\frac{1}{\zeta_\mathrm{CV}\tilde{\mu}_\mathrm{CV}}$.
	Hence, Theorem~1 is proved.
	\hfill \rule{4pt}{8pt}\\	
	
	Theorem~1 provides insights into the maintenance of content freshness.
	Firstly, the freshness of user received contents depends on the service rate of both VM-1 and VM-2, indicating that the greedy RSU cache update scheme (i.e., maximizing the resource allocated to VM-1) may not be AoI-optimal.
	Secondly, the resource allocated to VM-1 should increase linearly with the density of publishers, to maintain the same-level of content freshness.
	
	\subsection{Service Delay Analysis} 
	
	\textbf{Theorem~2.} The average service delay of the CIS slice is
	\begin{equation}
	\label{eq_D_c}
	\bar{D}_\mathrm{C} = \frac{1}{\zeta_\mathrm{CV} \tilde{\mu}_\mathrm{CV}-2\pi J R_\mathrm{R} \lambda_\mathrm{C}}.
	\end{equation}
	
	\emph{Proof.}~For the CIS slice, the service delay corresponds to the dwelling at VM-2.
	The service process of VM-2 can be modeled as an M/M/1 queue, with arrival rate of $\Lambda_\mathrm{C}$ and service rate of ${\mu}_\mathrm{CV} = \zeta_\mathrm{CV} \tilde{\mu}_\mathrm{CV}$.
	Thus,
	\begin{equation}
	\bar{D}_\mathrm{C} = \frac{1}{\mu_\mathrm{CV}-\Lambda_\mathrm{C}} = \frac{1}{\zeta_\mathrm{CV} \tilde{\mu}_\mathrm{CV}-2\pi J R_\mathrm{R} \lambda_\mathrm{C}},
	\end{equation}
	and Theorem~2 is proved.
	\hfill \rule{4pt}{8pt}\\
	
	Notice that the service latency increases with the traffic load, requiring that the service rate of VM-2 should be increased in the heavily-loaded case.
	However, this can reduce the service rate of VM-1, degrading content freshness.
	This result indicates that there exist a AoI-delay tradeoff relationship in the CIS slice, influenced by the intra-slice resource allocation. 
	To achieve the AoI-delay balance, the weighted sum of average AoI and delay should be minimized by optimizing resource slicing ratio $\zeta_\mathrm{CS}$ and $\zeta_\mathrm{CV}$.
	As the average AoI and delay are both convex with respect to the resource slicing ratio (proved by taking second-order derivatives), this optimization problem is convex and can be easily solved by MATLAB toolbox.

\section{Infotainment Slice Management}
    \label{sec_II_analysis}
    
    This section analyzes the achievable throughput of the IS slice, including the local breakout traffic and the RSU unicast traffic.
    The throughput of local breakout depends on the on-board hit rate, which is the key issue to analyze.
    The achievable throughput of RSU unicast depends on the service rate of VM-4 and the availability of VM-3, showing coupling effect in the inter-slice resource allocation.
    
    \subsection{On-Board Hit Rate}
    
    \begin{figure}[!t]
    	\centering
    	\includegraphics[width=2.2in]{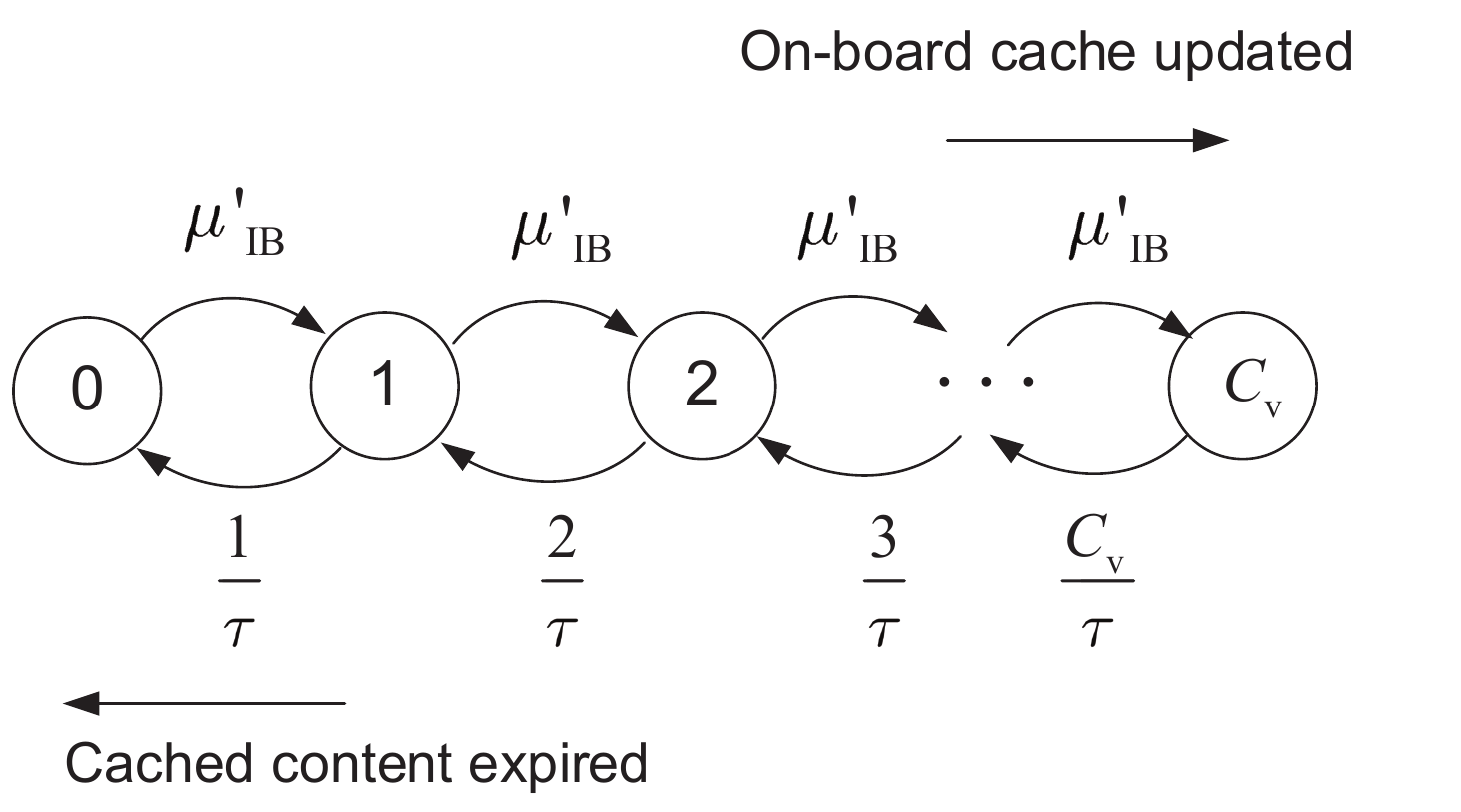}
    	\caption{The variation of the number of effective contents cached on-board.}
    	\label{fig_Mchain_I}
    \end{figure} 
    
    As the popularity of a content can decay with time in an exponential manner \cite{popularity_exponential_delay_conf_2007,popularity_exponential_delay_conf_2009}, VM-3 should keep broadcasting the newly generated popular contents to vehicles to guarantee the content hit rate.
    The challenge is the random expiration and update process.
    Denote by $c$ the number of effective content items cached on board, where $c\in[0,1,2,\cdots, C_\mathrm{v}]$.
    Assume that the lifetime of a content item follows exponential distribution of mean $\tau$, and the variation of $c$ can be modeled by a Markov chain.
    The state transition is illustrated as Fig.~\ref{fig_Mchain_I}, where $\mu'_\mathrm{IB}=\mu_\mathrm{IB}+\Delta_\mathrm{IB}$ is the aggregated broadcast rate by reusing VM-2.
    As the probability that VM-2 is in the idle state equals to
    \begin{equation}
    o_\mathrm{VM2} = 1-\frac{\Lambda_C}{\zeta_\mathrm{CV} \mu_\mathrm{CV}},
    \end{equation}
    $\Delta_\mathrm{IB}$ is given by 
    \begin{equation}
    \Delta_\mathrm{IB} = \zeta_\mathrm{CV} \tilde{\mu}_\mathrm{IB} - \Lambda_C,
    \end{equation}
    according to Eq.~(\ref{eq_delta_ib}).
    Denote by $\Phi=[\phi_1,\phi_2,\cdots, \phi_{C_\mathrm{v}}]$ the stable state probability distribution.
    The local balance function is given by
    \begin{equation}
    \label{eq_phi_local_balance}
    \frac{i}{\tau} \phi_i = \mu'_\mathrm{IB} \phi_{i-1},
    \end{equation}
    for $i=1,2,...,C_\mathrm{v}$.
    Substituting Eq.~(\ref{eq_phi_local_balance}) into $\sum_{i=0}^{C_\mathrm{v}} \phi_i = 1$ yields:
    \begin{equation}
    \label{eq_stable_P_I}
    \phi_0 = \frac{1}{\sum_{i=0}^{C_\mathrm{v}}\frac{(\tau \mu'_\mathrm{IB})^i}{i!}},~~~\mbox{and}~~~
    \phi_i = \frac{\frac{(\tau \mu'_\mathrm{IB})^i}{i!}}{\sum_{i=0}^{C_\mathrm{v}}\frac{(\tau \mu'_\mathrm{IB})^i}{i!}}.
    \end{equation}
    Notice that the stable state probability distribution is a Poisson distribution truncated on the right.
    The content hit rate is:
    \begin{equation}
    \label{eq_hit_rate}
    P_\mathrm{Hit} = \sum_{i=1}^{C_\mathrm{v}} \phi_i \sum_{f=1}^{i} q_f  = \frac{\sum_{i=1}^{C_\mathrm{V}} \sum_{f=1}^{i} \frac{(\tau \mu'_\mathrm{IB})^i}{i!} q_f }{\sum_{i=0}^{C_\mathrm{v}}\frac{(\tau \mu'_\mathrm{IB})^i}{i!}} .
    \end{equation}
    In specific, we study two typical content popularity distributions (i.e., uniform and Zipf) to analyze the key influencing factors on the on-board hit rate.
    
    \textbf{Proposition~1.} If the popularity of IS contents follows uniform distribution, the on-board content hit rate has closed-form expression:
    \begin{equation}
    {P}_\mathrm{Hit} = \frac{C_\mathrm{v}}{F} \cdot \frac{\tau\mu'_\mathrm{IB}\Gamma(C_\mathrm{v},\tau\mu'_\mathrm{IB})}{\Gamma(C_\mathrm{v}+1,\tau\mu'_\mathrm{IB})},
    \end{equation}
    where $\Gamma(\cdot,\cdot)$ is the upper incomplete gamma function. 
    The on-board hit rate increases with $\tau\mu'_\mathrm{IB}$ and converges to $C_\mathrm{v}/F$ as $\tau\mu'_\mathrm{IB}\rightarrow\infty$.\\
    
    \emph{Proof.} If all files are requested with equal probability, $q_f = \frac{1}{F}$ and then Eq.~(\ref{eq_hit_rate}) can be rewritten as
    \begin{subequations}
    	\label{eq_P_hit_uniform}
    	\begin{align}
    	{P}_\mathrm{Hit} & = \frac{\sum_{i=1}^{C_\mathrm{v}} \frac{(\tau \mu'_\mathrm{IB})^i}{i!} \cdot \frac{i}{F}}{\sum_{i=0}^{C_\mathrm{v}} \frac{(\tau \mu'_\mathrm{IB})^i}{i!}}\\
    	& = \frac{\tau\mu'_\mathrm{IB}}{F}\frac{\sum_{i=0}^{C_\mathrm{v}-1} \frac{(\tau \mu'_\mathrm{IB})^i}{i!} \cdot e^{-\tau\mu'_\mathrm{IB}}}{\sum_{i=0}^{C_\mathrm{v}}\frac{(\tau \mu'_\mathrm{IB})^i}{i!} \cdot e^{-\tau\mu'_\mathrm{IB}}}\\
    	&= \frac{\tau\mu'_\mathrm{IB}}{F} \frac{\frac{\Gamma(C_\mathrm{v},\tau\mu'_\mathrm{IB})}{(C_\mathrm{v}-1)!}}{\frac{\Gamma(C_\mathrm{v}+1,\tau\mu'_\mathrm{IB})}{C_\mathrm{v}!}}\\
    	& = \frac{\tau\mu'_\mathrm{IB}C_\mathrm{v}}{F} \frac{\Gamma(C_\mathrm{v},\tau\mu'_\mathrm{IB})}{\Gamma(C_\mathrm{v}+1,\tau\mu'_\mathrm{IB})},
    	\end{align}
    \end{subequations}
    where (\ref{eq_P_hit_uniform}b) and (\ref{eq_P_hit_uniform}c) come from the cumulated distribution function of Poisson distributions.
    We can prove that $P_\mathrm{Hit}$ increases with $\tau \mu'_\mathrm{IB}$ by taking derivative of $P_\mathrm{Hit}$, which is intuitive and omitted due to the page limit.
    In addition,
    \begin{equation}
    \label{eq_hit_rate_limit_norm}
    \begin{split}
    \lim\limits_{\tau\mu'_\mathrm{IB}\rightarrow \infty} P_\mathrm{Hit} & = \lim\limits_{\tau\mu'_\mathrm{IB}\rightarrow \infty} \frac{\sum_{i=1}^{C_\mathrm{v}} \frac{(\tau \mu'_\mathrm{IB})^i}{i!} \cdot \frac{i}{F}}{\sum_{i=1}^{C_\mathrm{v}} \frac{(\tau \mu'_\mathrm{IB})^i}{i!}} \\
    & =  \lim\limits_{\tau\mu'_\mathrm{IB}\rightarrow \infty} \frac{\frac{(\tau \mu'_\mathrm{IB})^{C_\mathrm{v}}}{C_\mathrm{v}!} \cdot \frac{C_\mathrm{v}}{F} }{\frac{(\tau \mu'_\mathrm{IB})^{C_\mathrm{v}}}{C_\mathrm{v}!}} = \frac{C_\mathrm{v}}{F}.
    \end{split}
    \end{equation}
    Hence, Proposition~1 is proved.
    \hfill \rule{4pt}{8pt}
    
    Proposition 1 indicates the content hit rate is mainly influenced by cache size $C_\mathrm{v}$ and $\tau \mu'_\mathrm{IB}$.
    $\tau \mu'_\mathrm{IB}$ can be defined as the \emph{normalized update rate}, reflecting the ratio of new content update rate to old content expire rate.
    Furthermore, $\lim\limits_{\tau\mu'_\mathrm{IB}\rightarrow \infty} P_\mathrm{Hit} = C_\mathrm{v}/F$ suggests that the hit rate depends on how much content can be cached on-board if the communication resource is not constraint.
    In this case, the on-board cache is always full of effective contents, since the popular contents are pushed shortly once generated. \\
    
	\textbf{Proposition~2.} If the popularity of IS contents follows Zipf distribution of parameter $\nu$, the on-board content hit rate has explicit-form expression:
    \begin{equation}
    {P}_\mathrm{Hit} =  \frac{\sum_{i=1}^{C_\mathrm{v}}\frac{(\tau\mu'_\mathrm{IB})^i}{i!}H_{i,\nu}}{\sum_{i=0}^{C_\mathrm{v}}\frac{(\tau\mu'_\mathrm{IB})^i}{i!}H_{F,\nu}},
    \end{equation}
    where $H_{n,\nu}$ is the $n$th generalized harmonic number:
    \begin{equation}
    H_{n,\nu} = \sum_{k=1}^{n} \frac{1}{k^\nu},
    \end{equation}
    and 
    \begin{equation}
    \label{eq_hit_rate_limit_zipf}
    \lim\limits_{\tau\mu'_\mathrm{IB}\rightarrow\infty} P_\mathrm{Hit} = \frac{H_{C_\mathrm{v},\nu}}{H_{F,\nu}}.
    \end{equation}\\
    
    \emph{Proof.} If the popularity of IS contents follows Zipf distribution, we have 
    \begin{equation}
    \label{eq_zipf}
    q_f = \frac{1/f^\nu}{\sum_{j=1}^{F} 1/j^\nu},
    \end{equation}
    where $\nu$ is the skewness factor reflecting the degree of request concentration.
    In specific, larger $\nu$ means that content requests show higher similarity.
    The typical value is $\nu=0.56$, corresponding to the video-type infotainment services \cite{Gill07_youtube}.
    Substitute Eq.~(\ref{eq_zipf}) into Eq.~(\ref{eq_hit_rate}), and we have	
    \begin{subequations}
    	\label{eq_P_hit_Zipf}
    	\begin{align}
    	{P}_\mathrm{Hit} & = \frac{\sum_{i=1}^{C_\mathrm{v}} \frac{(\tau \mu'_\mathrm{IB})^i}{i!} \cdot \sum_{f=1}^{i}\frac{1}{f^{\nu}}}{\sum_{i=0}^{C_\mathrm{v}} \frac{(\tau \mu'_\mathrm{IB})^i}{i!} \sum_{j=1}^{F}\frac{1}{j^{\nu}}}\\
    	& = \frac{\sum_{i=1}^{C_\mathrm{v}}\frac{(\tau\mu'_\mathrm{IB})^i}{i!}H_{i,\nu}}{\sum_{i=0}^{C_\mathrm{v}}\frac{(\tau\mu'_\mathrm{IB})^i}{i!}H_{F,\nu}},
    	\end{align}
    \end{subequations}
    where (\ref{eq_P_hit_Zipf}b) comes from the cumulated distribution function of Zipf's law.
    Eq.~(\ref{eq_hit_rate_limit_zipf}) can be proved in the same way as Eq.~(\ref{eq_hit_rate_limit_norm}).
    \hfill \rule{4pt}{8pt}\\
    
    Similar to the uniform popularity case, the on-board hit rate is mainly constrained by the content update rate and cache size, whereas the trend and performance is different.
    For instance, the on-board hit rate increases with cache size $C_\mathrm{v}$ and $\nu$ according to Eq.~(\ref{eq_hit_rate_limit_zipf}).
    Thus, the on-board hit rate is larger if the content requests are more concentrated.
    Note that the uniform popularity distribution can be considered as a special case of Zipf popularity by taking $\nu=0$.
    Therefore, proactive content pushing is more beneficial in case of Zipf distribution, especially for larger $\nu$.
    
    \subsection{Local Breakout}
    
    \textbf{Theorem 3.} The probability of local breakout in the IS slice is given by
    \begin{equation}
    \label{eq_P_LB_expl}
    P_\mathrm{LB} = \frac{\sum_{i=1}^{C_\mathrm{V}} \sum_{f=1}^{i} \frac{(\tau \mu'_\mathrm{IB})^i}{i!} q_f }{\sum_{i=0}^{C_\mathrm{v}}\frac{(\tau \mu'_\mathrm{IB})^i}{i!}} \left(1 - (1-P_\mathrm{vc}) e^{-\pi R_\mathrm{v}^2 P_\mathrm{vc} \rho_\mathrm{v}} \right),
    \end{equation}
    where $R_\mathrm{v}$ is the transmission range of V2V communications.
    
    \emph{Proof.}  	
    The probability of local breakout is given by (\ref{eq_P_LB}), where the on-board hit rate $P_\mathrm{Hit}$ has been obtained as (\ref{eq_hit_rate}), and the remaining issue is to analyze $P_\mathrm{V2V}$, i.e., the probability that a vehicle can be assisted by surrounding vehicles.
    As the spatial distribution of vehicles follows one-dimensional PPP of density $\rho_\mathrm{v}$ in each lane, the distribution of cache-enabled vehicles also follows PPP of density $P_\mathrm{vc}\rho_\mathrm{v}$.
    Consider a vehicle with no cache instance.
    The number of cache-enabled vehicles within the V2V communication range follows Poisson distribution of $2 J \pi R_\mathrm{v} P_\mathrm{vc} \rho_\mathrm{v}$.
    Accordingly, $P_\mathrm{V2V}$ can be obtained:
    \begin{equation}
    \label{eq_P_v2v}
    P_\mathrm{V2V} = 1 - e^{-2 J \pi R_\mathrm{v} P_\mathrm{vc} \rho_\mathrm{v}},
    \end{equation}
    which is the probability that there is at least one cache-enabled vehicle in the V2V communication range.
    Substitute (\ref{eq_hit_rate}) and (\ref{eq_P_v2v}) into (\ref{eq_P_LB}), the local breakout probability is obtained as (\ref{eq_P_LB_expl}).
    \hfill \rule{4pt}{8pt}\\
    
    In (\ref{eq_P_LB_expl}), the first part corresponds to the probability that the requested content is stored on cache-enabled vehicles, and the second part reflects the probability that a vehicle is cache-enabled or has cache-enabled vehicles within V2V communication range.
    Notice that the first part increases with both the cache size $C_\mathrm{v}$ and broadcast rate $\mu'_\mathrm{IB}$.
    The second part increases with the cache-enabled probability $P_\mathrm{vc}$, the vehicle density, and the V2V communication range, all in concave manner.
    In general, more traffic can be served locally with more cache-enabled vehicles' assistance.
    
    \subsection{Throughput of Infotainment Slice}
    
    The aggregated throughput of IS slice is given by
    \begin{equation}
    U_\mathrm{I} = \min\left\{\Lambda_\mathrm{I}, \Lambda_\mathrm{I} P_\mathrm{LB} + \zeta_\mathrm{IU} \tilde{\mu}_\mathrm{IU} + \zeta_\mathrm{IB}\tilde{\mu}_\mathrm{IU} \phi_{C_\mathrm{v}}\right\},
    \end{equation}
    where $\Lambda_\mathrm{I} P_\mathrm{LB}$ is the throughput of local breakout, $\zeta_\mathrm{IU} \tilde{\mu}_\mathrm{IU}$ is the achievable rate of RSU unicast with VM-4, and $ \zeta_\mathrm{IB}\tilde{\mu}_\mathrm{IU} \phi_{C_\mathrm{v}}$ corresponds to the multiplexing gain by reusing VM-3.
    $\phi_{C_\mathrm{v}}$ is the probability that the on-board cache is full of effective contents and VM-3 is in the idle state.
    Maximizing the IS throughput is non-intuitive considering the coupling effect.
    Increasing broadcast resource can increase on-board hit rate and the probability of local breakout $P_\mathrm{LB}$, as well as the idle probability of VM-3 $\phi_{C_\mathrm{v}}$.
    However, the service rate of VM-4 can be significantly reduced.
    In practice, the numerical results of optimal intra-slice resource allocation ratio can be obtained by searching methods, based on the derived analytical results.
    
\section{Simulation and Numerical Results}
    \label{sec_simulation}
    
    This section conducts simulations to evaluate the analytical results of both slices, using OMNeT++ and Matlab simulators.
    In addition, the multiplexing gain of soft slicing is shown, compared with the hard slicing without resource reuse.
    Furthermore, the multi-dimensional QoS performance of both slices is demonstrated under the proposed soft slicing scheme, which provides insights into how to meet the requirements of differentiated applications.
    Important simulation parameters are listed in Table~\ref{tab_parameter}.
    
    \begin{table}[!t]
    	\caption{Simulation parameters}
    	\label{tab_parameter}
    	\centering
    	\begin{tabular}{cccc}
    		\hline
    		\hline
    		Parameter & Value & Parameter & Value \\
    		\hline
    		$B$ & 10 MHz & $L$ & 1 Mb \\
    		$P_\mathrm{TP}$ & 0.1 W & $P_\mathrm{TR}$ & 1 W \\
    		$\sigma^2$ & -105 dBm	& $I_\mathrm{Intf}$ & -75 dBm \\
    		$\alpha$ & 3.5 & $J$ & 4 \\
    		$R_\mathrm{R}$ & 400 m	& $R_\mathrm{v}$  & 100 m \\
    		$\rho_\mathrm{v}$ & 5 /km	& $\rho_\mathrm{p}$ & 1000 /km$^2$ \\
    		$F$ & 1000 & $C_\mathrm{v}$ & 10 \\
    		$P_\mathrm{vc}$ & 0.5 & $\tau$ & 30 s\\
    		\hline
    		\hline
    	\end{tabular}
    \end{table}

    \begin{figure*}[!t]
    	\centering
    	\subfloat[] {\includegraphics[width=2.5in]{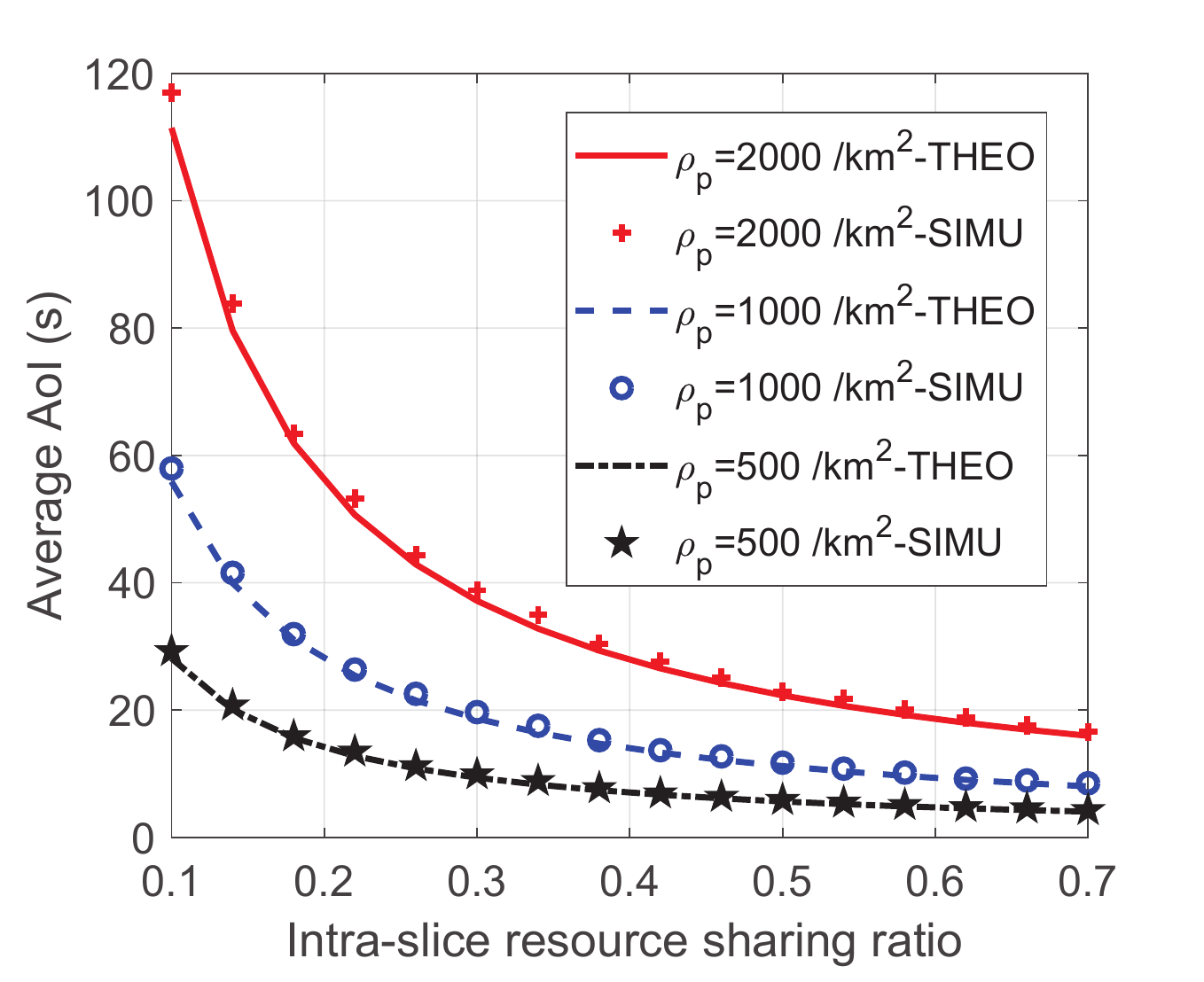}}
    	\subfloat[]{\includegraphics[width=2.5in]{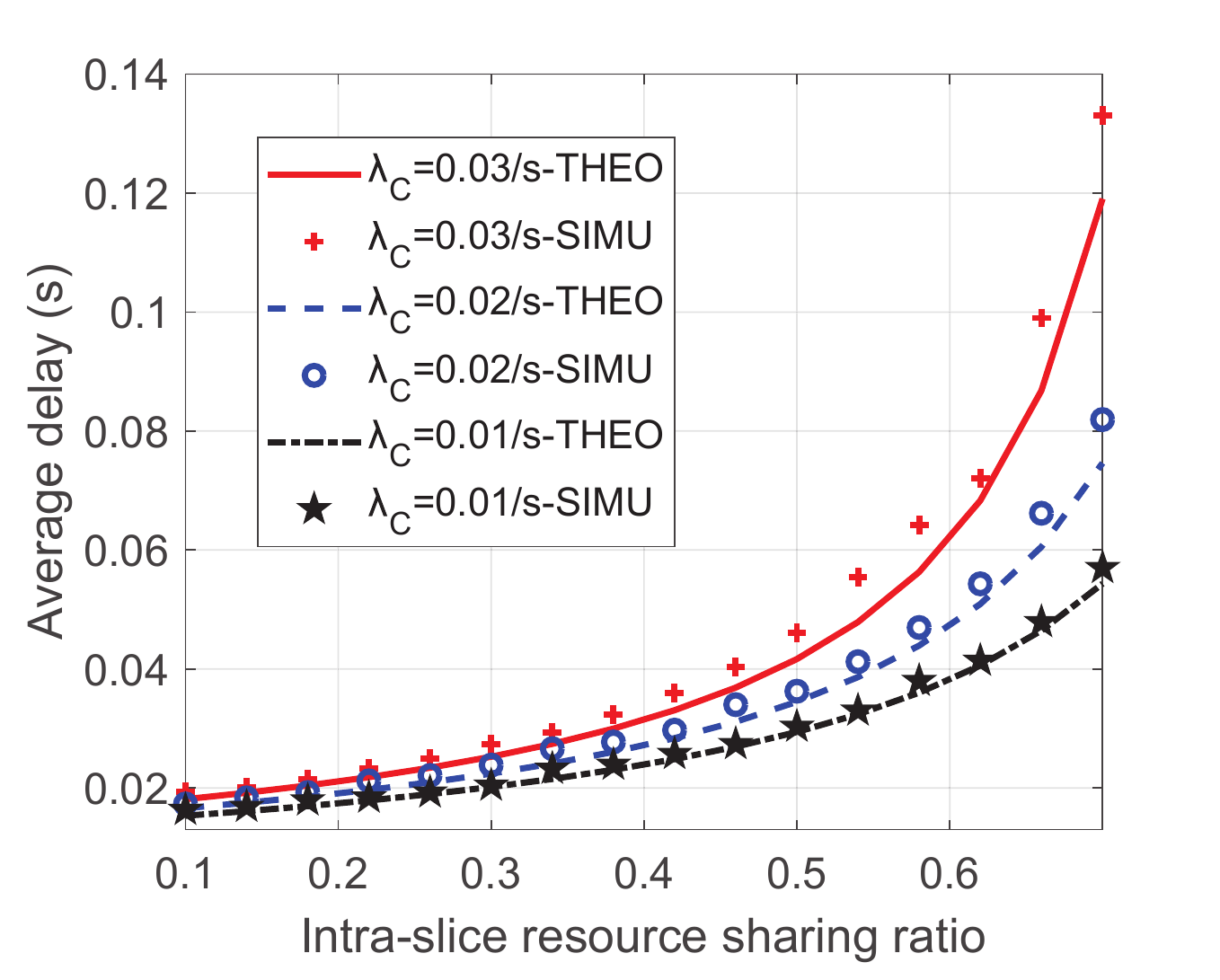}}
    	\caption{Analytical result evaluation of CIS slice, (a) average AoI ($\lambda_\mathrm{C} = 0.02$ /s), and (b) average delay ($\rho_\mathrm{p}=1000$ /km$^2$).}
    	\label{fig_CIS_evaluation}
    \end{figure*}
    
    \begin{figure*}[!t]
    	\centering
    	\subfloat[] {\includegraphics[width=2.5in]{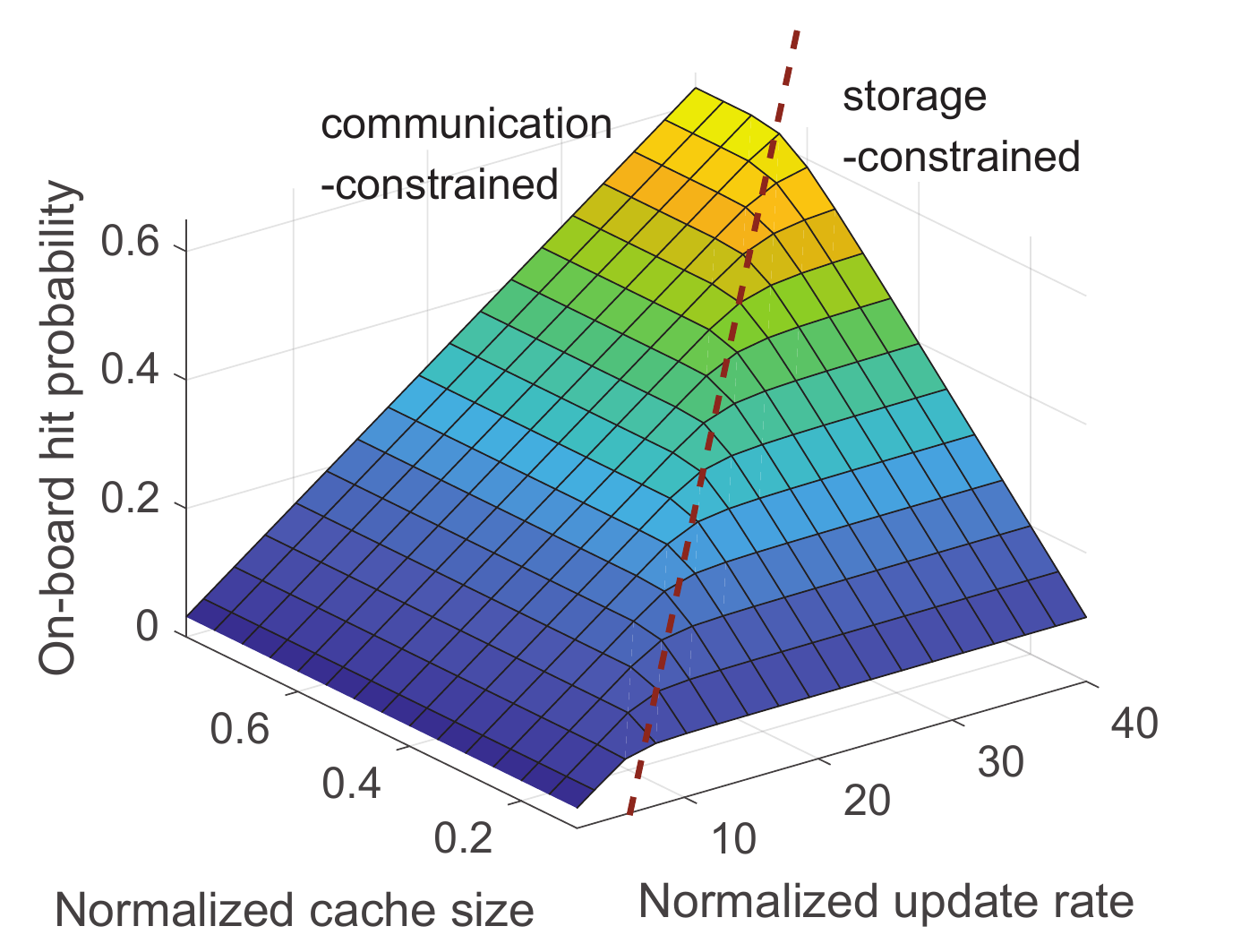}}
    	\subfloat[]{\includegraphics[width=2.5in]{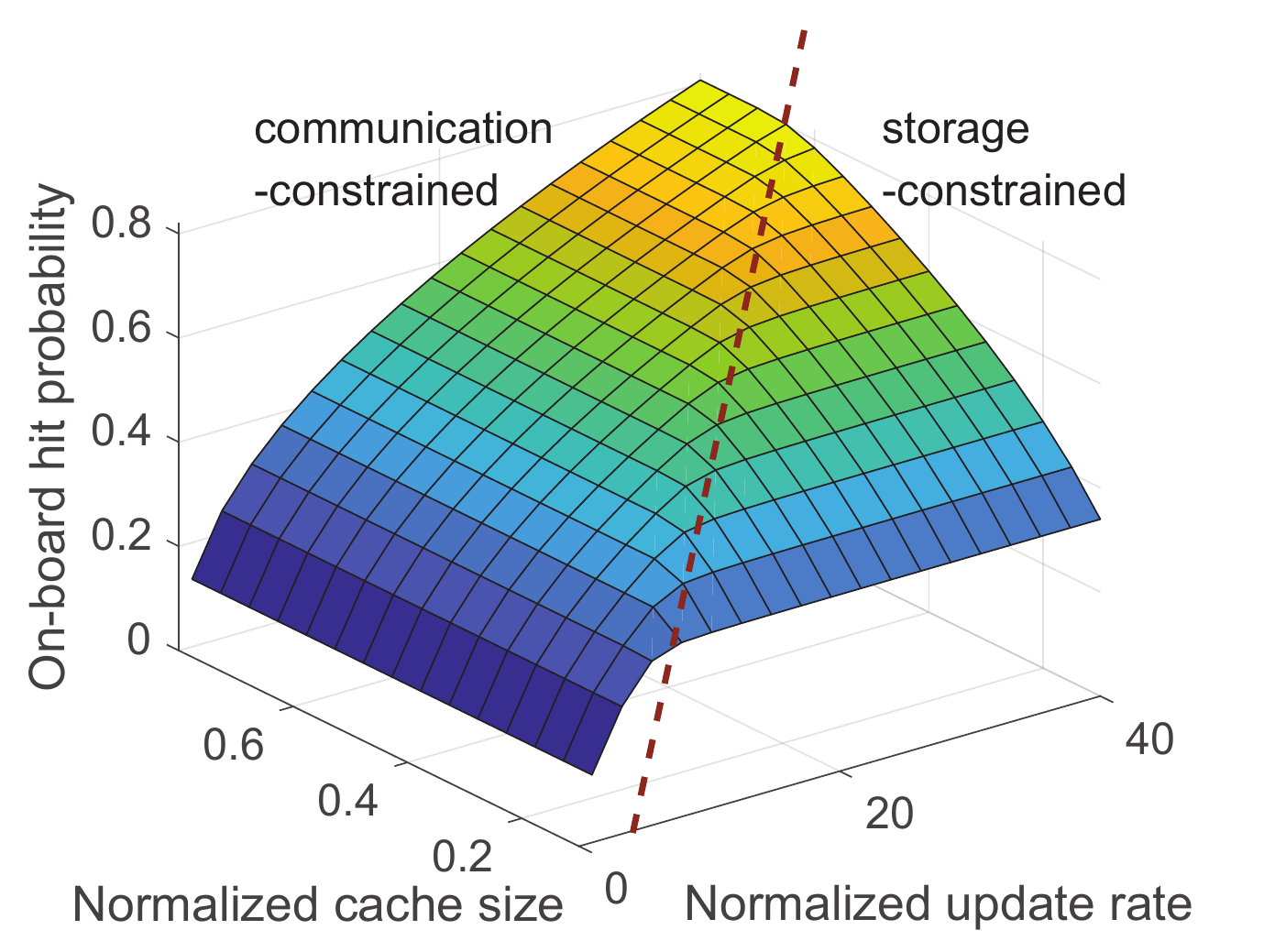}}
    	\caption{On-board hit rate of the IS contents, (a) uniform content popularity, and (b) Zipf content popularity.}
    	\label{fig_onboard_hit_rate}
    \end{figure*}

    \subsection{Context Information Slice Analysis}
    
    For the CIS slice, we validate the derived average AoI and delay by conducting system level Monte Carlo simulations on the OMNeT++ platform.
    All bandwidth is allocated to the CIS slice, whereas the intra-slice resource allocation ratio is tuned to meet different AoI and delay requirements.
    The number of vehicles, their locations and requests are all generated randomly, whereby the AoI and delay of each request can be obtained. 
    Then the average AoI and delay are calculated based on the 1000 simulations, as shown in Fig.~\ref{fig_CIS_evaluation}.
    The analytical results are obtained by Eqs.~(\ref{eq_A_v}) and (\ref{eq_D_c}).
    Figure~\ref{fig_CIS_evaluation} shows that the derived average AoI and delay are quite close to the simulation under different parameter settings, which validates the theoretical analysis.
    With more resources allocated to VM-1, the average AoI decreases while the average delay increases.
    This result indicates that the intra-slice resource allocation ratio should be adjusted to satisfy the specific requirement of applications in practice.
    Furthermore, the average AoI is shown to increase significantly with the publisher density, indicating higher costs to maintain content freshness.
    Instead, the average delay is shown to rely on the request rate.
    These results are consistent with the Theorems~1 and 2.

    \subsection{Infotainment Slice Analysis}
    
    For the IS slice, the numerical results of on-board hit rate, local breakout probability, and the achievable throughput are illustrated, showing the key influencing system parameters including resource slicing ratio, content popularity distribution, vehicle density, cache size, and cache-enabled ratio.
    Only the IS slice is considered, with all bandwidth allocated.
    
    The on-board hit rate is shown in Figs.~\ref{fig_onboard_hit_rate}(a) and \ref{fig_onboard_hit_rate}(b), with respect to the cache size normalized by the file library (i.e., $C_\mathrm{v}/F$) and the normalized update rate (i.e., $\tau \mu'_\mathrm{IB}$).
    The results show that the on-board hit rate increases with both cache size and update rate, which is consistent with the analytical results.
    Furthermore, for the given cache size, the on-board hit rate firstly increases with the update rate but then levels off.
    In specific, the on-board hit rate is shown to converge to the normalized cache size for the uniform content popularity case, which validates Proposition~1.
    For instance, the on-board hit rate is 0.1497 when the vehicle can cache 15\% of all files, if the content update rate is 40 files/s.
    In this case, the update rate is sufficiently large to push the newly generated popular contents to vehicles instantly, and thus the on-board cache is always full of effective contents.
    Instead, the on-board hit rate is mainly constrained by the on-board cache size.
    Therefore, the on-board hit rate can be constrained either by the communication source or the storage resource, as demonstrated in Fig.~\ref{fig_onboard_hit_rate}(a).
    The results also show that larger cache size requires higher update rate to achieve the maximal hit rate, revealing that the storage resource and communication resource should be balanced for efficient utilization.
    In case of the Zipf content popularity distribution, we can draw the similar conclusions from the numerical results, which validates Proposition~2.
    In comparison, the on-board hit rate is shown to be higher under the Zipf content popularity distribution, in both the communication-constrained and storage-constrained regions. 
    This is because the cached popular contents are requested more frequently, increasing on-board hit rate.
    
    \begin{figure}[!t]
    	\centering
    	\includegraphics[width=2.5in]{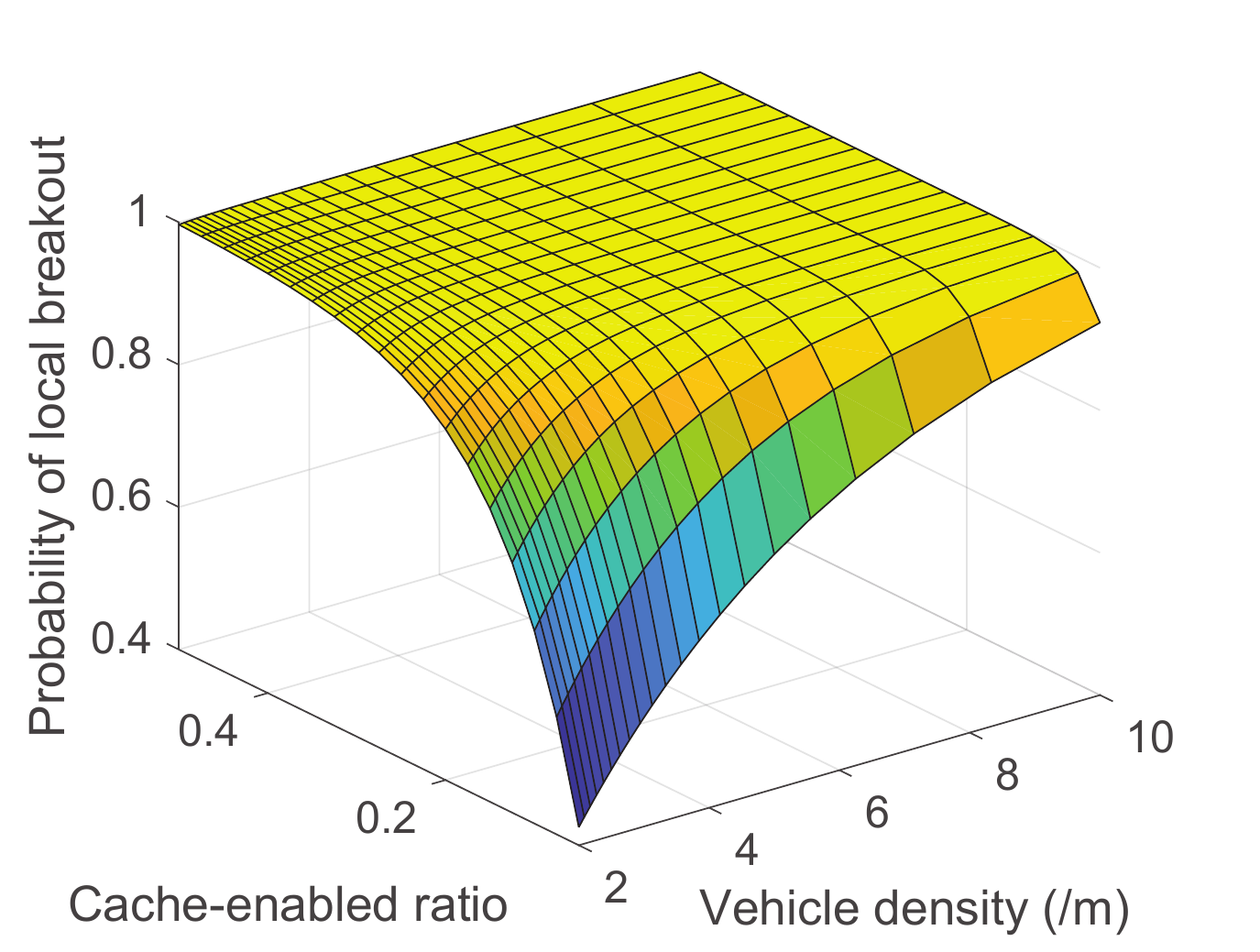}
    	\caption{Local breakout probability, normalized by on-board hit rate.}
    	\label{fig_local_breakout}
    \end{figure}
    
    \begin{figure*}[!t]
    	\centering
    	\subfloat[] {\includegraphics[width=2.5in]{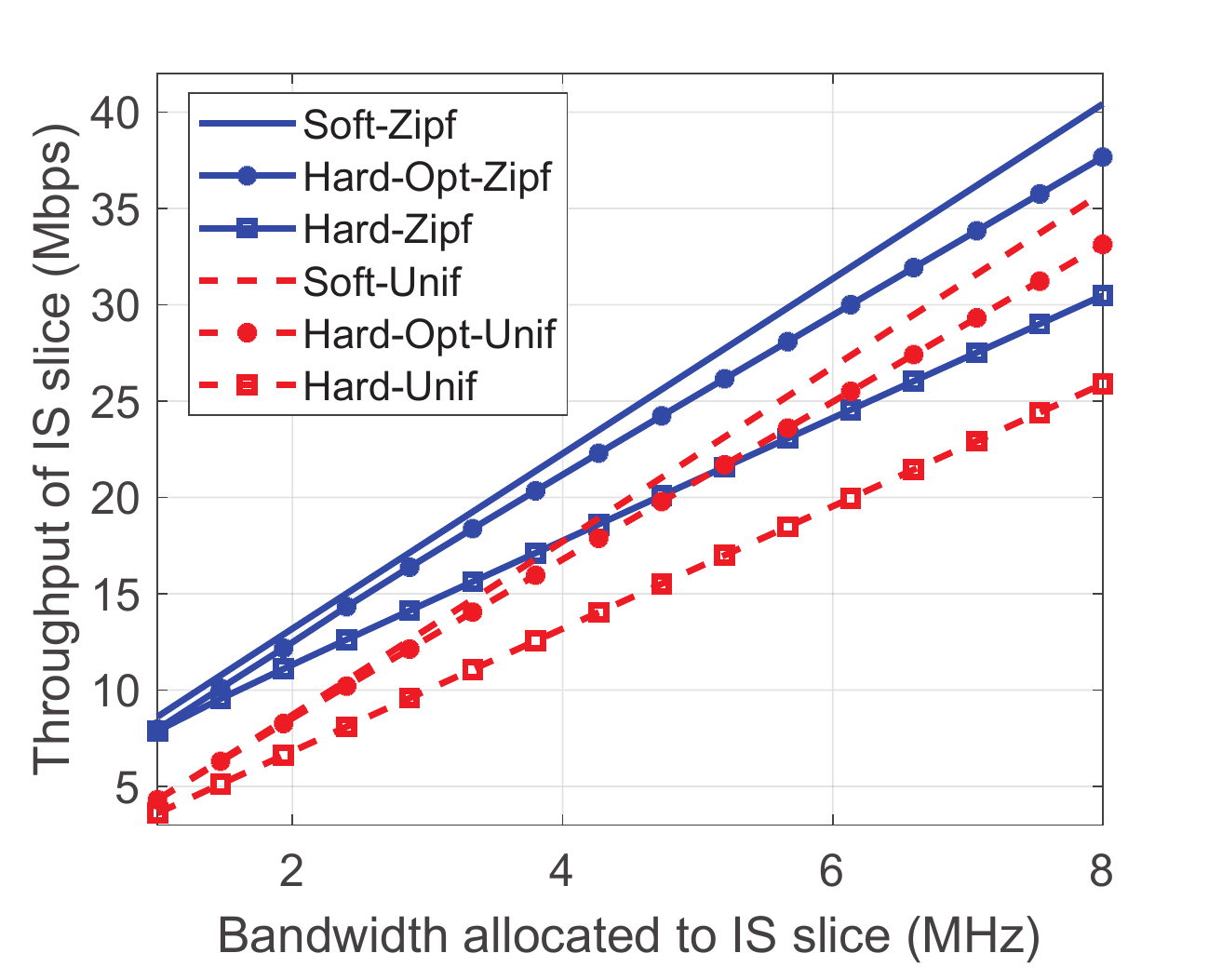}}
    	\subfloat[]{\includegraphics[width=2.5in]{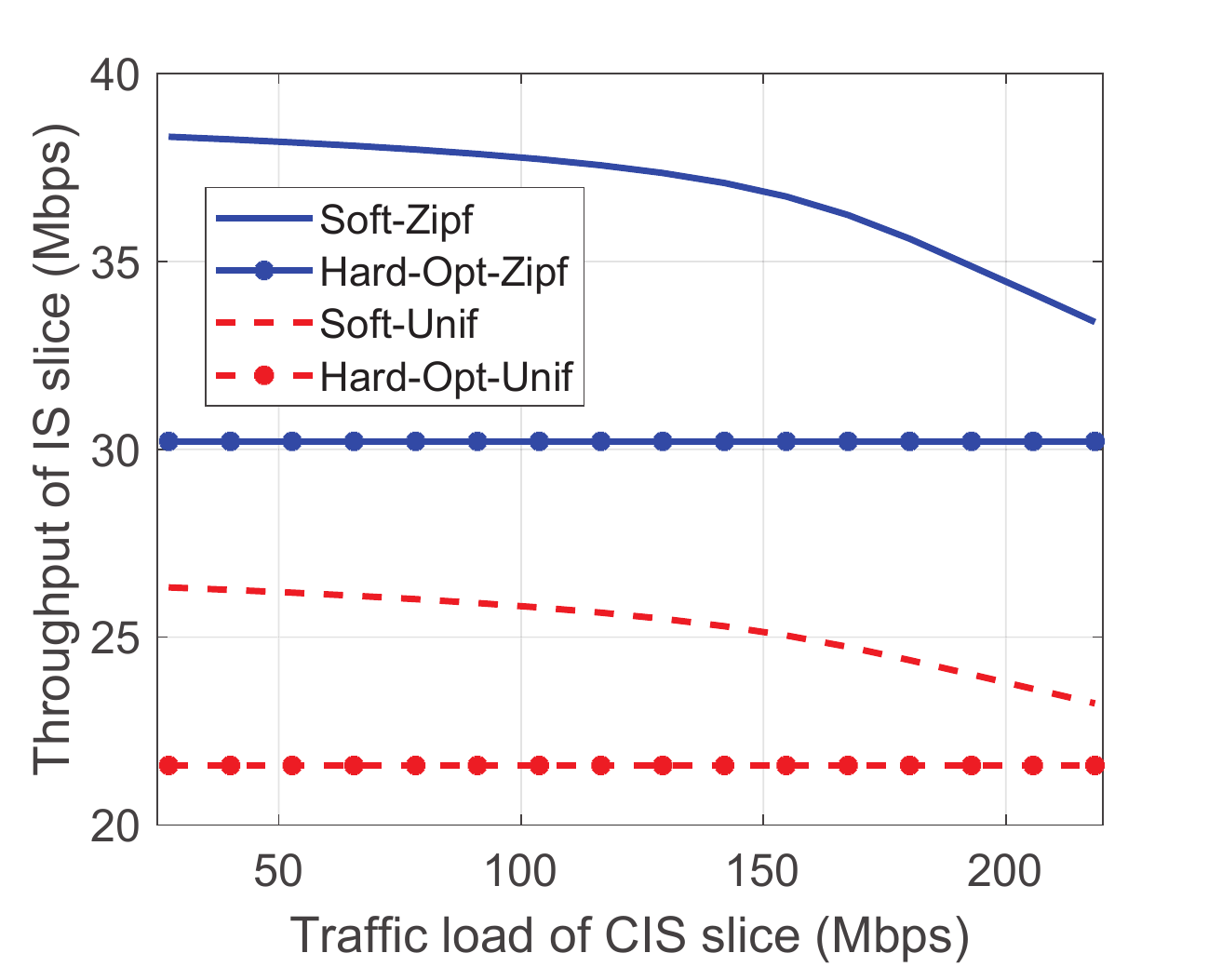}}
    	\caption{IS throughput gain through soft slicing, (a) gain of intra-slice resource reuse, and (b) gain of inter-slice resource reuse.}
    	\label{fig_throughput_IS}
    \end{figure*}

    \begin{figure}[!t]
    	\centering
    	\subfloat[] {\includegraphics[width=1.8in]{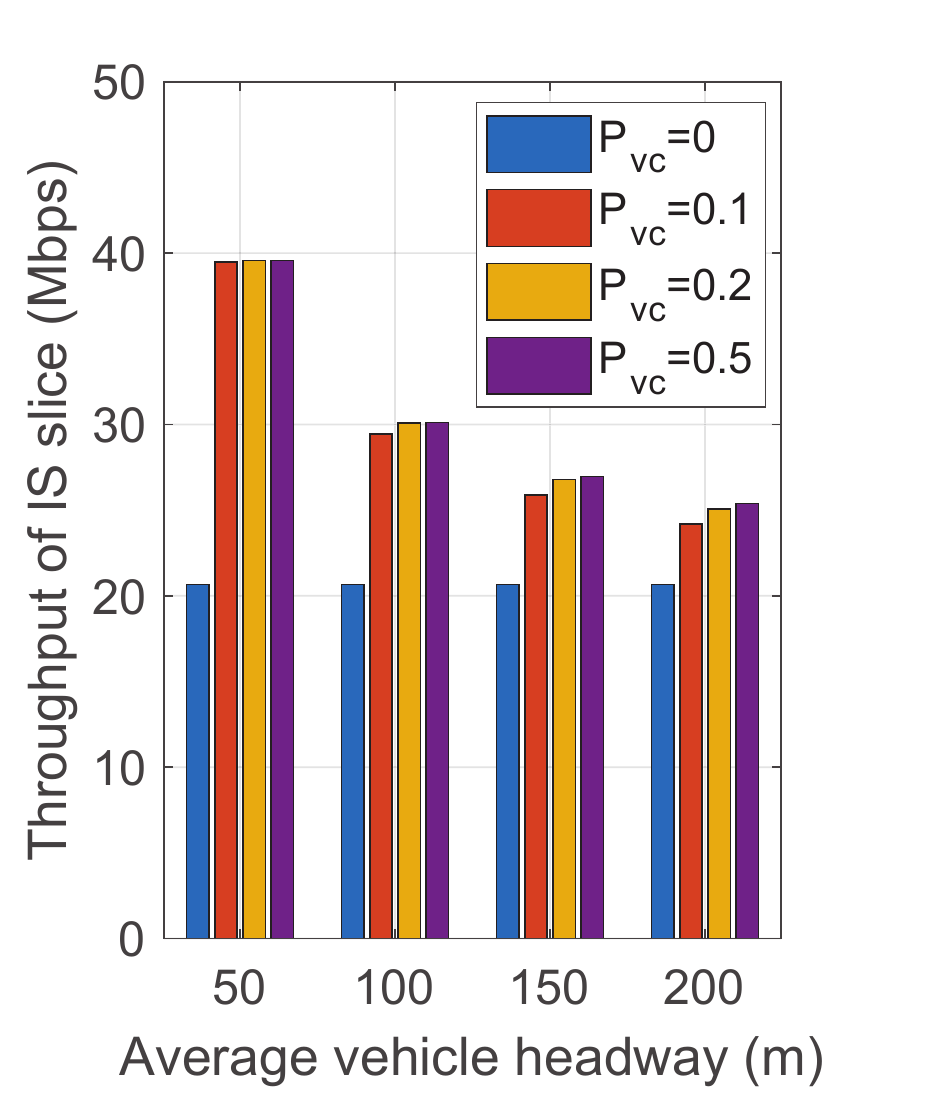}}
    	\subfloat[]{\includegraphics[width=1.8in]{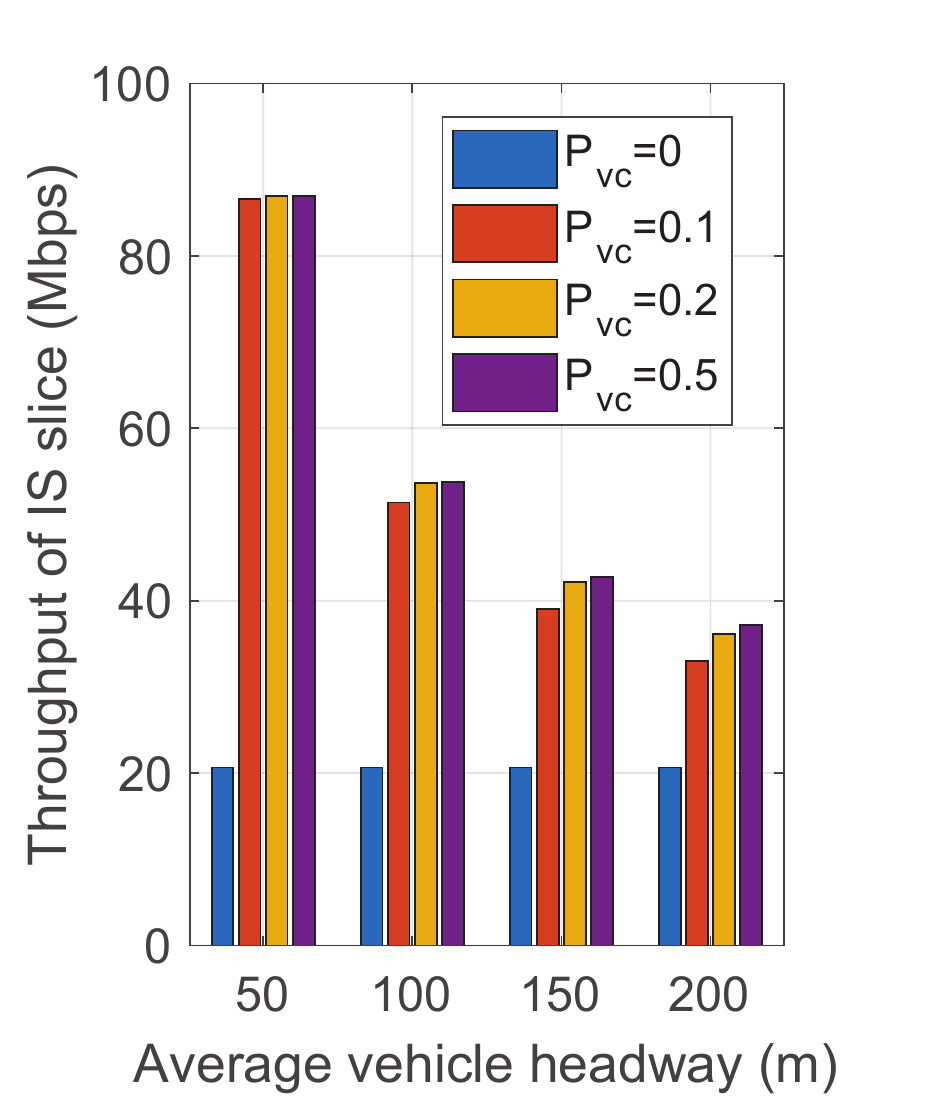}}
    	\caption{Influencing factors on IS throughput, (a) Uniform content popularity distribution, and (b) Zipf content popularity distribution.}
    	\label{fig_throughput_IS_bar}
    \end{figure}
    
    Figure~\ref{fig_local_breakout} shows the probability of local breakout which is normalized by on-board content hit rate, with respect to the cache-enabled ratio (i.e., $P_\mathrm{vc}$) and vehicle density (i.e., $\rho_\mathrm{v}$).
    The physical meaning is the probability that a vehicle is either cache-enabled or can find at least one cache-enabled vehicle within V2V communication range.
    The result of Fig.~\ref{fig_local_breakout} shows that the probability of local breakout increases with both vehicle density and cache-enabled ratio.
    When the vehicle density increases, a vehicle without cache has a higher chance to find a cache-enabled vehicle for V2V assistance.
    In addition, higher cache-enabled ratio indicates that more vehicles are equipped with on-board cache, and thus vehicles are more likely to realize local breakout.
    When the normalized probability of local breakout reaches 1, it is equivalent that all vehicles are equipped with on-board cache.
    In this case, the achievable throughput of local breakout mainly depends on the on-board content hit rate.

    \begin{figure*}[!t]
    	\centering
    	\subfloat[] {\includegraphics[width=2.5in]{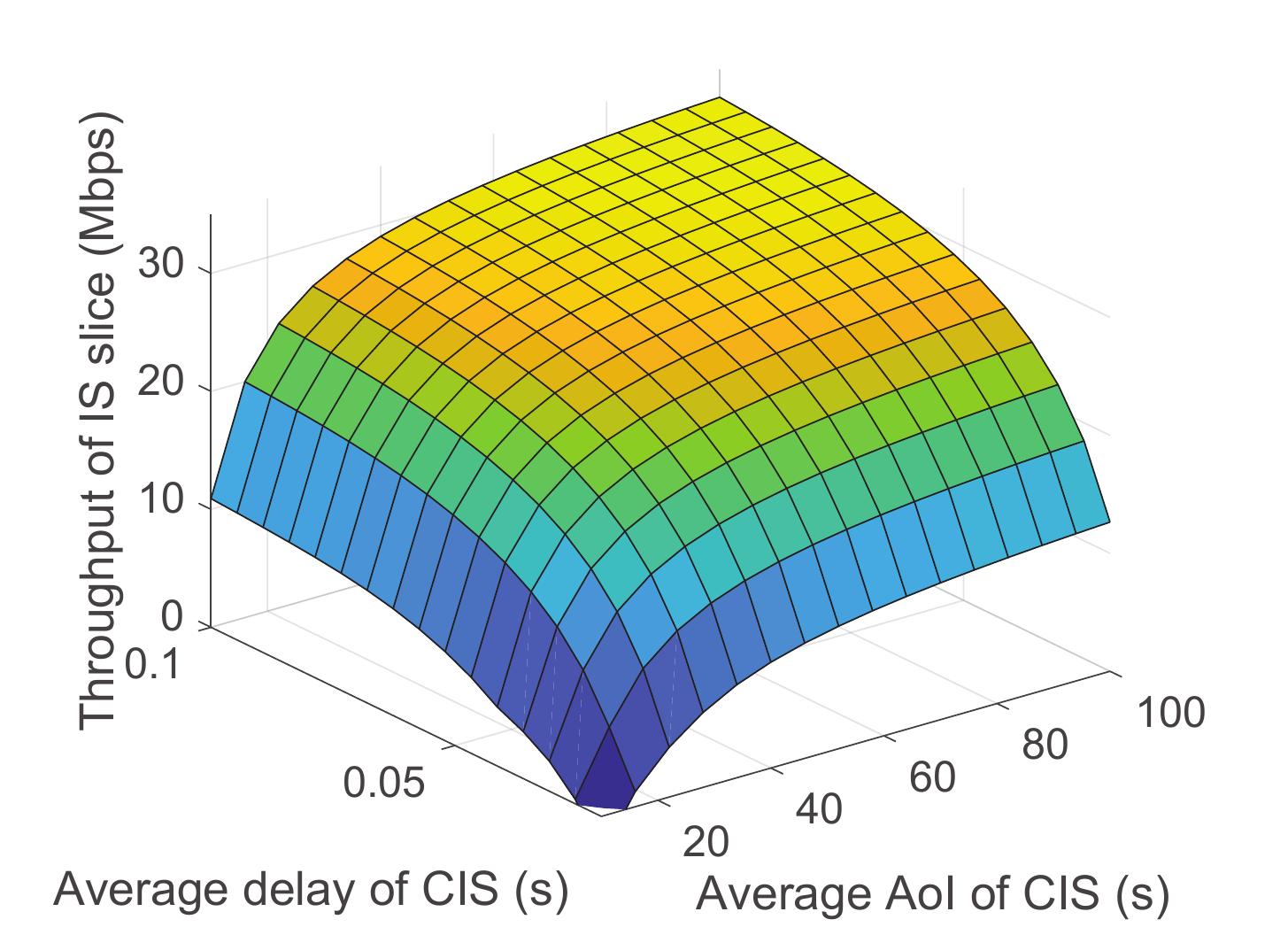}}
    	\subfloat[]{\includegraphics[width=2.5in]{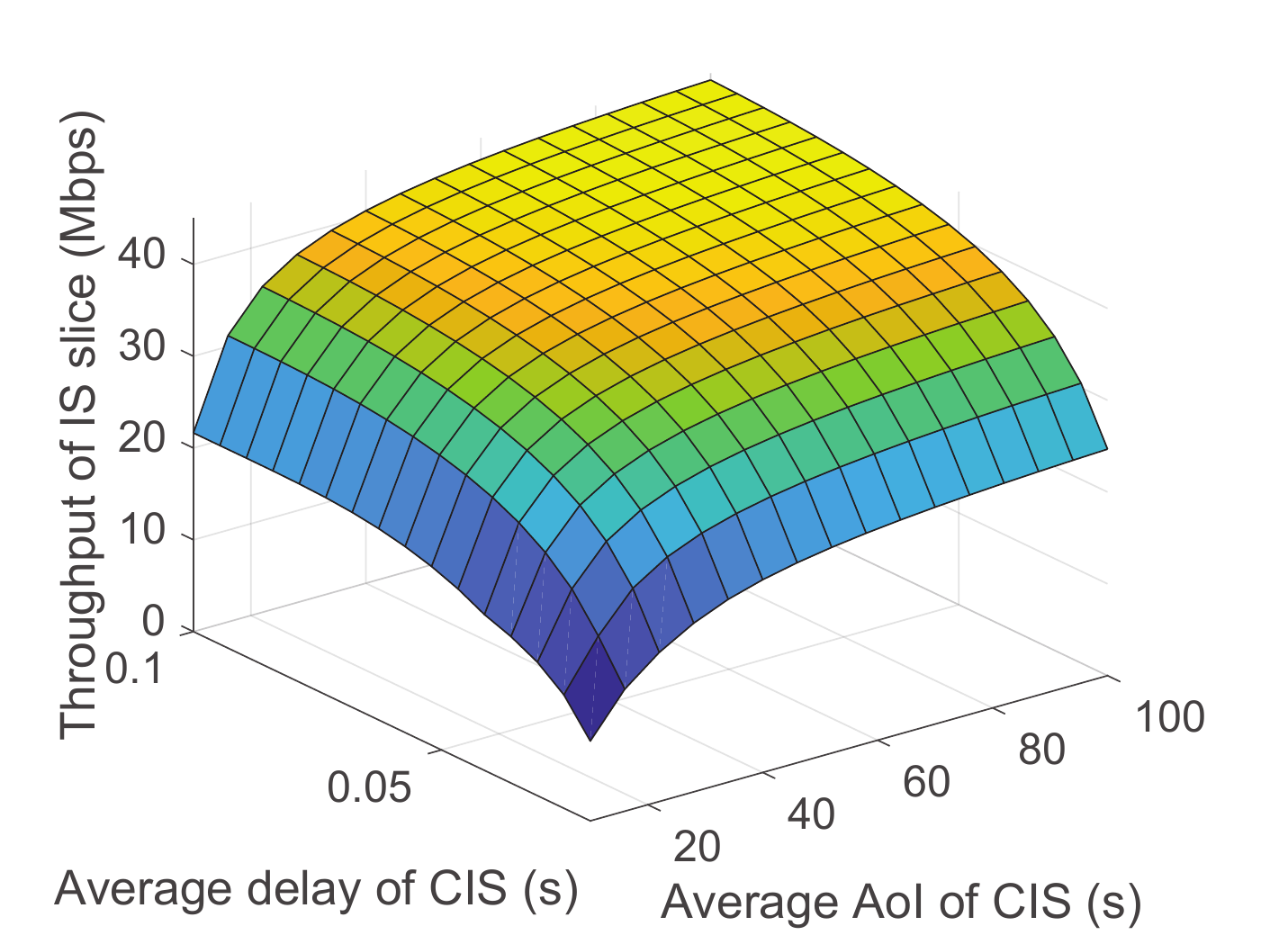}}
    	\caption{Three-dimensional QoS performance of CIS and IS slices of soft slicing, (a) uniform content popularity distribution, and (b) Zipf content popularity distribution.}
    	\label{fig_QoS_3D}
    \end{figure*}
    
    \subsection{Multiplexing Gain}
    
    With the proposed soft slicing method, the resources can be utilized opportunistically in the intra- and inter-slice manners, which can enhance the throughput of the IS slice without influencing the performance of the CIS slice.
    To evaluate the performance of soft slicing, we compare the throughput of the IS slice under both soft and hard slicing schemes, as shown in Fig.~\ref{fig_throughput_IS}.
    Figure~\ref{fig_throughput_IS}(a) shows the throughput gain obtained by intra-slice resource reuse in the IS slice, where the resource allocated for RSU pro-active content push (i.e., VM-3) can be reused by RSU unicast content delivery (i.e., VM-4).
    Two hard slicing schemes without resource reuse are adopted as baselines: (1) constant hard slicing scheme, whereby 30\% IS bandwidth is allocated to RSU pro-active content push and 70\% IS bandwidth is allocated to RSU unicast content delivery; and (2) optimal hard slicing, where the resource allocated to RSU pro-active content push and unicast content delivery is optimized.
    The results of Fig.~\ref{fig_throughput_IS}(a) show that the throughput of IS slice can be improved by enabling opportunistic intra-slice resource reuse, which can bring around 30\% and 10\% throughput gain compared with the constant and optimal hard slicing schemes, respectively.
    Figure~\ref{fig_throughput_IS}(b) further shows the throughput gain through inter-slice resource reuse, whereby the idle resource of CIS slice can be reused by the IS slice.
    The bandwidth allocation ratio is set to $\zeta_\mathrm{CS}=0.15$, $\zeta_\mathrm{CV}=0.35$, $\zeta_\mathrm{IB}+\zeta_\mathrm{IU}=0.5$ while $\zeta_\mathrm{IB}$ and $\zeta_\mathrm{IU}$ are further optimized under both schemes to maximize the throughput.
    Compared with the optimal hard slicing, the throughput can be improved by around 30\% compared with the optimal soft slicing scheme, when the CIS load is low.
    In this case, more CIS resources are available for inter-slice reuse, and the proposed soft slicing method is more beneficial.
    
    The throughput of IS slice is also affected by system parameters, as shown in Fig.~\ref{fig_throughput_IS_bar}.
    $P_\mathrm{vc}=0$ corresponds to the case that no vehicle is cache-enabled and all traffic is served through RSU unicast.
    The results of Fig.~\ref{fig_throughput_IS_bar}(b) indicate that the throughput can be improved by more than threefold through local breakout if 10\% vehicles are cache-enabled.
    In addition, the throughput can be further enhanced as the ratio of cache-enabled vehicles increases, whereas the marginal effect is significant.
    The important insight is that equipping partial vehicles (such as taxis and buses) with cache instances can be great helpful to enhance the IS capacity of vehicular networks.
    Furthermore, the throughput is shown to increase with vehicle density (i.e., decrease with average vehicle headway), as more content delivery requests are raised in this case.

    \subsection{Three-Dimensional QoS Guarantee}

    The bandwidth demand of CIS slice can be obtained for the given average AoI and delay requirements according to Eqs.~(\ref{eq_A_v}) and (\ref{eq_D_c}), determining the inter-slice resource allocation.
    Then, the intra-slice resource allocation of IS slice can be optimized for maximum throughput.
    In this way, the achievable three-dimensional QoS performance of soft slicing can be obtained, as shown in Fig.~\ref{fig_QoS_3D}.
    In specific, the result demonstrates a three-dimensional tradeoff relationship.
    For example, increasing the IS throughput needs to sacrifice either the average AoI or the delay performance of the CIS slice.
    In addition, the average AoI and delay can be traded in the CIS slice, for the given IS throughput requirement.
    The main reason of this tradeoff relationship is the resource constraint.

\section{Conclusions and Future Work}
    \label{sec_conclusions}

	This paper has proposed a hierarchical soft slicing framework for cache-enabled vehicular networks supporting the typical CIS and IS applications, where the RSU resource is reused opportunistically within and across slices.
	In specific, the average AoI and delay of the CIS slice have been derived, showing a tradeoff relationship with respect to intra-slice resource allocation.
	Furthermore, the throughput of the IS slice has been obtained with inter- and intra-slice opportunistic resource reuse, considering the local breakout through proactive content caching and sharing among vehicles.
	The analytical results have been validated through OMNeT++ simulations, and numerical results have been provided to reveal the interplay between the performance of different slices.
	In specific, the achievable QoS performance shows a three-dimensional tradeoff relationship, depending on the inter- and intra-slice resource sharing.
	In addition, the proposed soft slicing can effectively improve the throughput of IS slice while guaranteeing the AoI and delay performance of the CIS slice, compared with conventional hard slicing methods.	
	Future works will consider non-ideal environments (i.e., the intermittent V2V transmissions), more practical constraints (e.g., unequal file size), and more types of applications (i.e., multi-service slicing).



\bibliographystyle{IEEEtran}

\end{document}